\documentclass{acm_proc_article-sp}

\newtheorem{lemma}{Lemma}
\newtheorem{theorem}[lemma]{Theorem}
\newtheorem{corollary}[lemma]{Corollary}
\newtheorem{proposition}[lemma]{Proposition}
\newtheorem{definition}{Definition}

\def\ba{\begin{array}}
\def\ea{\end{array}}
\def\bi{\begin{itemize}}
\def\ei{\end{itemize}}
\def\bd{\begin{description}}
\def\ed{\end{description}}
\def\bu{\begin{enumerate}}
\def\eu{\end{enumerate}}
\def\be{\begin{equation}}
\def\ee{\end{equation}}
\def\bn{\begin{eqnarray}}
\def\en{\end{eqnarray}}

\def\bproof{\noindent{\bfseries\itshape Proof.}\hskip1pc\ignorespaces}
\def\endproof{{\hfill \Large $\bf\bigstrut \Box$}\vskip .15cm}

\def\bt{\begin{theorem}\rm}
\def\et{\end{theorem}}
\def\bc{\begin{corollary}\rm}
\def\ec{\end{corollary}}
\def\bl{\begin{lemma}\rm}
\def\el{\end{lemma}}
\def\bp{\begin{proposition}\rm}
\def\ep{\end{proposition}}
\def\bdf{\begin{definition}\rm}
\def\edf{\end{definition}}
\def\bx{\begin{example}}
\def\ex{\end{example}}

\def\ds{\displaystyle}

\def\qed{\hbox{\hfill []}}
\def\cal{\mathcal}

\def\A{{\bf A}}
\def\B{{\bf B}}
\def\C{{\bf C}}
\def\D{{\bf D}}
\def\E{{\bf E}}
\def\F{{\bf F}}

\def\K{{\bf K}}

\def\P{{\bf P}}
\def\Q{{\bf Q}}

\def\S{{\bf S}}

\def\V{{\bf V}}

\def\X{{\bf X}}
\def\Y{{\bf Y}}

\def\a{{\bf a}}
\def\b{{\bf b}}
\def\c{{\bf c}}
\def\d{{\bf d}}

\def\i{{\bf i}}

\def\l{{\bf l}}
\def\m{{\bf m}}
\def\n{{\bf n}}

\def\q{{\bf q}}

\def\s{{\bf s}}
\def\t{{\bf t}}
\def\u{{\bf u}}
\def\v{{\bf v}}
\def\w{{\bf w}}
\def\x{{\bf x}}
\def\y{{\bf y}}
\def\z{{\bf z}}

\def\0{{\bf 0}}
\def\1{{\bf 1}}
\def\2{{\bf 2}}
\def\3{{\bf 3}}
\def\4{{\bf 4}}
\def\5{{\bf 5}}
\def\6{{\bf 6}}
\def\7{{\bf 7}}
\def\8{{\bf 8}}
\def\9{{\bf 9}}

\def\bq{\bar{\bf q}}

\def\bQ{\bar{\bf Q}}

\def\is{{{\cal I}^\square}}

\def\lt{{\rm lt}}

\newbox\bigstrutbox
\setbox\bigstrutbox=\hbox{\vrule height12.5pt depth5pt width0pt}
\def\bigstrut{\relax\ifmmode\copy\bigstrutbox\else\unhcopy\bigstrutbox\fi}

\newbox\Bigstrutbox
\setbox\Bigstrutbox=\hbox{\vrule height17.5pt depth5pt width0pt}
\def\Bigstrut{\relax\ifmmode\copy\Bigstrutbox\else\unhcopy\Bigstrutbox\fi}

\setpagenumber{1}

\begin{document}

\title{Normalization of Polynomials in Algebraic Invariants of Three-Dimensional Orthogonal Geometry}

\numberofauthors{1}

\author{\alignauthor Hongbo Li\\ 
\vskip .1cm
       \affaddr{KLMM, AMSS, Chinese Academy of Sciences, Beijing 100190, China}\\
\vskip .1cm
       \email{\small
       hli@mmrc.iss.ac.cn}   
}

\maketitle

\begin{abstract} \vskip .1cm
In classical invariant theory, the Gr\"obner base of the ideal of syzygies and the normal forms of
polynomials of invariants are two core contents. To improve the performance of invariant theory in symbolic
computing of classical geometry, advanced invariants are introduced via Clifford product \cite{li07}.
This paper addresses and solves the two key problems in advanced invariant theory: 
the Gr\"obner base of the ideal of syzygies among advanced invariants, and the normal forms of
polynomials of advanced invariants. These results beautifully extend the straightening of 
Young tableaux to advanced invariants.
\end{abstract}


\section{Introduction}
\vskip .2cm

In traditional analytic approach to classical geometry, coordinates are introduced to represent
points in the geometric space, and equations of the coordinates are used to define constraints among
the points, forming a representation of higher dimensional objects such as curves, surfaces, etc.
Basic manipulations of coordinates include addition and multiplication, resulting in polynomials
in the coordinates. Since the coordinates of generic points are independent, and
the multiplication of coordinate variables are commutative, normalization of the polynomials
in the coordinates is very easy. The normal forms of the polynomials are required in many manipulations,
{\it e.g.}, division among polynomials.

Another analytic approach to classical geometry, dating back to Euclid, is to use geometric invariants 
such as lengths, angles, areas,
{\it etc}. A typically algebraic system of geometric invariants is a polynomial ring generated by basic invariants.
In such a system, a vector variable in a linear space
is used to represent a point or direction in classical geometry,
the inner product of a vector with itself represents the  
squared length of the vector, the inner product of two unit vectors represents the
cosine of the angle between them,
{\it etc}. Such operators among vectors generate a set of {\it basic invariants}, and the polynomials
in these basic invariants are {\it advanced invariants}. 

Although the multiplication of invariants 
are commutative, the basic invariants generated by generic vector variables of the linear space 
are not independent, and there are polynomial relations among them, called {\it syzygy} relations. 
The dependency is largely caused by
the dimension constraint of the linear space upon vectors. While the dimension constraint can be easily
reflected by the number of coordinates introduced to represent a point and the independency among the 
coordinates, for basic invariants generated by the points, fully representing
the dimension constraint is by no means trivial. Classical invariant theory studies the generators of
invariants, the syzygy relations
among the basic invariants, and the normal forms of advanced invariants as polynomials in 
the basic ones \cite{olver}, \cite{sturmfels}.

In symbolic geometric computing, both the coordinate approach and the basic invariant approach encounter
the difficulty of very big polynomial size, in particular in the middle of symbolic manipulations.
In \cite{li07}, a recipe to alleviate the difficulty is proposed, called long geometric product, 
BREEFS, and Clifford factorization, among which the long geometric product (or Clifford product)
is the foundation. The idea is to convert polynomials of basic invariants into advanced invariants,
converse to the approach of classical invariant theory, 
by means of an associative and multilinear product among the vector variables representing points. 
The associativity of the product and the symmetries within a long bracket provide powerful
manipulations that cannot be done with basic invariants, nor with coordinates. This is a 
top-down approach to advanced invariants \cite{li}, while the classical invariant theory is a bottom-up approach.

Dealing with the syzygy relations among advanced invariants and finding the normal forms of 
polynomials in advanced invariants are two fundamental tasks in such ``advanced invariant theory".
The 2D case is easy, while higher dimensional cases are difficult.
Little advance has been achieved in six years since the publication of \cite{li07} in 1997.

In this paper, the two fundamental problems are solved for the advanced invariant theory of 3D orthogonal
geometry: the Gr\"obner base of the syzygy ideal of ``long brackets", and the normal forms of Clifford
bracket polynomials. It turns out that the normal forms of such
bracket polynomials are surprisingly ``beautiful". The description is the following.

In classical invariant theory for $(n-1)$D projective geometry, the basic invariants are brackets of length
$n$, or in coordinate form, the $n\times n$ determinants formed by the homogeneous coordinates of $n$ 
vector variables. A bracket polynomial is in normal form if when each term is up to coefficient
written in Young tableau form, the entries in each row are increasing, while 
the entries in each column are non-decreasing \cite{white}. For example for vector variables
$\v_1\prec \v_2\prec \ldots \prec\v_m$, a bracket monomial 
$[\v_{i_{11}}\v_{i_{12}}\cdots \v_{i_{1c}}]
 [\v_{i_{21}}\v_{i_{22}}\cdots \v_{i_{2c}}]\cdots
[\v_{i_{r1}}\v_{i_{r2}}\cdots$ $ \v_{i_{rc}}]$, where the $\v_{i_{jk}}$ are repetitive selections
of the $m$ vector variables, is normal if and only if in
\be
\left[\ba{cccc}
\v_{i_{11}} & \v_{i_{12}} & \cdots & \v_{i_{1c}}\\
\v_{i_{21}} & \v_{i_{22}} & \cdots & \v_{i_{2c}}\\
\vdots & \vdots & \ddots & \vdots \\
\v_{i_{r1}} & \v_{i_{r2}} & \cdots & \v_{i_{rc}}
\ea\right],
\label{example:row}
\ee
$\v_{i_{j1}} \prec \v_{i_{j2}} \prec \cdots \prec \v_{i_{jc}}$, while
$\v_{i_{1k}} \preceq \v_{i_{2k}} \preceq \cdots \preceq \v_{i_{rk}}$.

In the advanced invariant theory for 3D orthogonal geometry, each ``elementary" advanced invariant is
a bracket of length $>1$, whose entries are vector variables representing 3D points. In a bracket 
monomial, different brackets many have different lengths, and a 
bracket monomial is in normal form if and only if not only the entries in each row are increasing, 
the entries in each column are non-decreasing, but all the entries in the tableau after removing the first
column, are non-decreasing.
For example if (\ref{example:row}) is normal in this setting, then the sequence
$\check{\v}_{i_{11}}\v_{i_{12}}\cdots \v_{i_{1c}}\check{\v}_{i_{21}}\v_{i_{22}} \cdots \v_{i_{2c}}
\cdots$ $\check{\v}_{i_{r1}}\v_{i_{r2}}\cdots \v_{i_{rc}}$ is non-decreasing, where 
$\check{\v}_{i_k}$ denotes that $\v_{i_k}$ does not occur in the sequence.

This paper is organized as follows. Section 2 introduces orthogonal geometric invariants by quaternions,
Clifford algebra and bracket algebra. Section 3 introduces the main results in \cite{li13} on
vector-variable polynomials and basics of ``advanced bracket algebra". Section 4 provides 
the Gr\"obner base and normal forms of long brackets in the multilinear case. Section 6 extends
the results to general case, by means of the square-free vector-variable polynomials introduced in
Section 5. Section 7 proposes a normalization algorithm for bracket polynomials.

\section{Quaternions, 
Clifford algebra, and bracket algebra}
\setcounter{equation}{0}
\vskip .2cm

In the vector algebra over ${\mathbb R}^3$, there are three multilinear products among vectors: 
(i) the inner product of two vectors, 
(ii) the cross product of two vectors, (iii) the hybrid product
of three vectors. None of them can be extended to include more vectors
while preserving the associativity.

The quaternionic product, on the other hand, is associative while still multilinear. Let $\bq$ represent the
quaternionic conjugate of quaternion $\q$. Among quaternions,
a {\it vector} $\v$ refers to a pure imaginary quaternion, {\it i.e.}, $\bar{\v}=-\v$,
and a {\it scalar} $\s$ refers to a real quaternion, {\it i.e.}, $\bar{\s}=\s$.
All vectors span a 3D real inner-product space
with metric diag$(-1,-1,-1)$, denoted by ${\mathbb R}^{-3}$.

We always use juxtaposition of elements to denote their quaternionic product.
The {\it inner product} of two vectors $\v_i, \v_j$
is defined by
\be
[\v_i\v_j]:=(\v_i\v_j+\v_j\v_i)/2. \label{inner}
\ee
The {\it cross product} of two vectors $\v_i, \v_j$
is defined by
\be
\v_i\times \v_j:=(\v_i\v_j-\v_j\v_i)/2.
\ee
The result is a vector, so its inner product with a third vector $\v_k$ is a scalar. Define
the {\it hybrid product} of three vectors $\v_i, \v_j, \v_k$ by
\be
[\v_i\v_j\v_k]:=(\v_i\v_j\v_k-\v_k\v_j\v_i)/2.
\ee
Then $[\v_i\v_j\v_k]=[(\v_i\times \v_j)\v_k]$. 

The vector algebra over ${\mathbb R}^{-3}$ is included in
the quaternions. The latter is equipped with a powerful product, 
the quaternionic product, making it possible
to use quaternions to represent 3D orthogonal transformations \cite{altmann}.

The {\it magnitude} of a quaternion $\q$ is $\sqrt{\overline{\q}\q}$.
A quaternion $\q$ is said to be {\it unit} if $\overline{\q}\q=1$.
Let $\q$ be a unit quaternion, and $\v$ be a vector. The {\it conjugate adjoint action} of $\q$ on $\v$
is defined by
\be
Ad_\q(\v):=\overline{\q}\v\q.
\ee
Since $[Ad_\q(\v_1)Ad_\q(\v_2)]=[\v_1\v_2]$ for any two vectors $\v_1,\v_2$, $Ad_\q$ realizes an orthogonal
transformation in ${\mathbb R}^{-3}$.
A classical result states that in fact all orthogonal
transformations in ${\mathbb R}^{-3}$ are realized in this way, and two different unit quaternions 
realize the same
orthogonal transformation if and only if they differ by sign.

For a quaternion $\Q$, the {\it bracket} $[\Q]$ is its scalar part:
\be
[\Q]:=(\Q+\bQ)/2.
\ee
The {\it axis} of $\Q$ is the vector part of $\Q$:
\be
A(\Q):=(\Q-\bQ)/2.
\ee
In particular, $A(\v_1\v_2)=\v_1\times \v_2$.
We interpret them in geometrical terms below.

For a unit vector $\v_1$, $Ad_{\v_1}$ realizes the reflection with respect to the plane normal to $\v_1$.
In general, for unit vectors $\v_1,\v_2,\ldots, \v_{2k+1}$, 
$Ad_{\v_1\v_2\cdots \v_{2k+1}}$ realizes the reflection with respect to the plane normal to axis
$A(\v_1\v_2\cdots \v_{2k+1})$, if the latter is nonzero.

For
two unit vectors $\v_1,\v_2$ that are linearly independent, 
$Ad_{\v_1\v_2}$ realizes the rotation about the axis $\v_1\times \v_2$: in the plane spanned by 
$\v_1, \v_2$, the rotation is from $\v_1$ to the reflection of $\v_1$ with respect to $\v_2$, {\it i.e.},
the angle of rotation is $\theta=2\angle (\v_1,\v_2)$. Furthermore, $[\v_1\v_2]=\cos(\theta/2)$.
When we say ``rotation $\v_1\v_2$", we mean the one induced by $Ad_{\v_1\v_2}$.

In general, for unit vectors $\v_1,\v_2,\ldots, \v_{2k}$, 
$Ad_{\v_1\v_2\cdots \v_{2k}}$ realizes the rotation about the axis  
$A(\v_1\v_2\cdots \v_{2k})$, if the latter is nonzero. The rotation is the composition of
$k$ rotations $\v_1\v_2, \v_3\v_4, \ldots, \v_{2k-1}\v_{2k}$. If the angle of rotation is 
$\theta$, then 
\be
{[}\v_1\cdots \v_{2k}] = \cos(\theta/2),\ \
|A(\v_1\cdots \v_{2k})| = |\sin(\theta/2)|.
\label{interpret:1}
\ee

Let $A(\v_1\v_2\cdots \v_{2k})\neq 0$. By
$[\v_1\v_2\cdots \v_{2k+1}] = [A(\v_1\v_2$ $\cdots \v_{2k})\v_{2k+1}]$, we get
\be
[\v_1\v_2\cdots \v_{2k+1}]=
\cos\angle(A(\v_1\v_2\cdots \v_{2k}),\v_{2k+1})\,
\sin(\theta/2),
\label{interpret:2}
\ee
where $\theta$ is the angle of rotation $\v_1\v_2\cdots \v_{2k}$.
In particular when $k=1$, for linearly independent unit vectors $\v_1, \v_2$, 
$\sin(\theta/2)=\sin\angle(\v_1, \v_2)$ equals the area of the parallelogram spanned by $\v_1, \v_2$,
and $\cos\angle(\v_1\times \v_2,\v_{3})$ equals the height from the end of unit vector
$\v_3$ to the plane spanned by $\v_1,\v_2$, so $[\v_1\v_2\v_3]$ equals the volume of the parallelepiped
spanned by $\v_1, \v_2, \v_3$.

In classical invariant theory, an {\it algebraic invariant} is a polynomial whose variables
are basic invariants. In 3D orthogonal geometry, there are two kinds of basic invariants: 
$[\v_i\v_j]$ and $[\v_i\v_j\v_k]$ for all vector variables $\v_i, \v_j, \v_k$.
Given $n$ vector variables $\v_1, \ldots, \v_n$, the brackets 
$[\v_{j_1}\v_{j_2}\cdots \v_{j_m}]$ for arbitrary $1<m<\infty$ and 
arbitrary repetitive selection of elements $\v_{j_1}, \v_{j_2}, \ldots, \v_{j_m}$ from the $n$ variables,
form an infinite set of advanced algebraic invariants. That each of them is a polynomial of the
$[\v_i\v_j]$ and $[\v_i\v_j\v_k]$ is guaranteed by the following {\it
Caianiello expansion formulas} \cite{brini}, \cite{li}: let $\V_k=\v_{j_1}\v_{j_2}\cdots \v_{j_k}$, then
\be\hskip -.2cm
\ba{cll}
{[}\V_{2l}] & \hskip -.2cm=& \hskip -.2cm\sum_{i=2}^{2l} 
(-1)^i [\v_{j_1}\v_{j_i}][\v_{j_2}\v_{j_3}\cdots \check{\v}_{j_i}\cdots \v_{j_{2l}}];\\

A(\V_{2l-1}) &\hskip -.2cm=& \hskip -.2cm \sum_{(2l-2,1)\vdash \V_{2l-1}} [{\V_{2l+1}}_{(1)}] {\V_{2l-1}}_{(2)}; \\

A(\V_{2l}) &\hskip -.2cm=& \hskip -.2cm \sum_{(2l-2,2)\vdash \V_{2l}} [{\V_{2l}}_{(1)}]A({\V_{2l}}_{(2)});\\

{[}\V_{2l+1}] &\hskip -.2cm=& \hskip -.2cm \sum_{(2l-2,3)\vdash \V_{2l+1}} [{\V_{2l+1}}_{(1)}][{\V_{2l+1}}_{(2)}],
\ea
\label{eq4}
\ee
where (i) 
$(h,m-h)\vdash \V_{m}$ is a bipartition of the $m$ elements in the sequence $\V_m$ into two subsequences
${\V_{m}}_{(1)}$ and ${\V_{m}}_{(2)}$ of length $h$ and $m-h$ respectively; (ii) in $[{\V_{m}}_{(1)}]$, the product of
the $h$ elements in the subsequence is denoted by the same symbol ${\V_{m}}_{(1)}$; (iii) the summation
$\sum_{(h,m-h)\vdash \V_{m}}$ is over all such bipartitions of $\V_m$, and the sign of permutation of
the new sequence ${\V_{m}}_{(1)}, {\V_{m}}_{(2)}$ is assumed to be carried by the first factor $[{\V_{m}}_{(1)}]$ 
of the addend. 

While quaternions are sufficient for describing orthogonal transformations in 3D, they cannot be generalized
to higher dimensions directly. In quaternions, the hybrid product $[\v_i\v_j\v_k]$ is a scalar. To make
high-dimensional generalization this requirement must be removed, at the same time the property that
this element be in the center of the algebra needs to be preserved. If we denote the quaternions by
$\cal Q$, then the above revision leads to a new algebra ${\cal Q}\oplus \iota{\cal Q}$ of dimension 8,
where $\iota:=[\v_1\v_2\v_3]$ for three fixed vector variables that are linearly independent.
This algebra is the {\it Clifford algebra} over ${\mathbb R}^{-3}$.

The formal definition of the {\it Clifford algebra} Cl$({\cal V}^n)$ over an $n$-dimensional 
$\mathbb K$-linear space ${\cal V}^n$, where the characteristic of $\mathbb K$ is $\neq 2$,
is the quotient of the tensor algebra $\bigotimes {\cal V}^n$ over the ideal generated by 
elements of the form $\v\otimes \v-Q(\v)$ where $Q$ is a $\mathbb K$-quadratic form.
The product induced from the tensor product is called the {\it Clifford product}, also denoted by 
juxtaposition of elements \cite{hestenes}, \cite{lounesto}. 

When ${\cal V}^n={\mathbb R}^{-3}$, the quaternionic product of vectors is the image of their
Clifford product under the homomorphism induced by mapping $\iota$ to a nonzero scalar. 
In Clifford algebra, $\iota$ is not a scalar, but called a {\it pseudoscalar} because it not only
commutes with everything, but spans a 1D real space containing all hybrid products.
The concept quaternionic conjugate is replaced by the {\it Clifford conjugate}, which is the linear extension
of the following operation: for any vectors  $\v_{j_1}, \v_{j_2}, \ldots, \v_{j_k}$, let
$\V_k=\v_{j_1}\v_{j_2}\cdots \v_{j_k}$, then
\be\ba{ll}
\overline{\V_k} := (-1)^k \V_k^\dagger,  \hbox{ where} &
\V_k^\dagger:=\v_{i_k}\cdots \v_{i_2}\v_{i_1}\\
& \hbox{ is the {\it reversion} of $\V_k$.}
\ea
\ee

With the Clifford conjugate, we can define the {\it magnitude} of $\V_k$, 
the {\it conjugate adjoint action} $Ad_{\V_k}$, 
the {\it bracket} $[\V_k]$,   
the {\it axis} $A(\V_k)$, together with the concepts not involving conjugate: 
the inner product $[\v_i\v_j]$, the cross product $\v_i\times \v_j$
and the hybrid product $[\v_i\v_j\v_k]$,
just the same as in the case of quaternions.
The only difference is that since $[\v_i\v_j\v_k]$ is now a pseudoscalar, while $A(\V_{2l-1})$ 
remains a vector, $A(\V_{2l})$ is not, but a {\it pseudovector}. Geometrically, when
$A(\v_1\v_2)\neq 0$, it represents the plane spanned by vectors $\v_1, \v_2$, or equivalently,
the invariant plane of rotation $Ad_{\v_1\v_2}$ supporting $\v_1, \v_2$.

We see that unlike the quaternions where there are only two kinds of objects of different dimensions:
scalars which are usually called 0-D objects, and vectors which represent 1-D 
directions and so are usually called 1-D objects,
in Clifford algebra Cl$({\mathbb R}^{-3})$ there are four kinds of objects of different dimensions.
Besides scalars and vectors, there are pseudovectors which represent 2-D 
directions (planes), and pseudoscalars which represent 3-D orientations (spaces). This is the
reason why Cl$({\mathbb R}^{-3})$ can be extended to higher dimensions by being capable of
discerning objects of different dimensions.

To represent algebraic invariants in 3D orthogonal geometry, using the quaternionic product or the Clifford product
in the brackets does not make any difference. The geometric interpretations (\ref{interpret:1}),
(\ref{interpret:2}) and the Caianiello expansion formulas are identical for both products.
This justifies the use of juxtaposition of elements to represent both products.

The two kinds of basic invariants $[\v_i\v_j]$ and $[\v_i\v_j\v_k]$ form a commutative ring, called
{\it inner-product bracket algebra}. Formally, given a set of $n$ symbols 
${\cal M}=\{\v_1, \ldots, \v_n\}$, two kinds of new symbols
can be defined as following: (1) all 2-tuples of symbols selected repetitively from $\cal M$, 
by requiring that each 2-tuple be symmetric with respect to its two entries; such a 2-tuple is denoted by
$[\v_i\v_j]$. (2) All 3-tuples of symbols selected repetitively from $\cal M$, 
by requiring that each 3-tuple is anti-symmetric with respect to its three entries,
and in particular, if there are identical entries in a 3-tuple, setting the 3-tuple to be zero;
such a 3-tuple is denoted by $[\v_i\v_j\v_k]$. 

The two kinds of symbols must satisfy the following \\
{\it dimension-three constraints}:
\bd
\item[$\rm IGP:$] For any five symbols $\v_{i_1}, \ldots, \v_{i_5}$,
\be\hskip -.9cm
\ba{lll}
\phantom{-} [\v_{i_1}\v_{i_2}][\v_{i_3}\v_{i_4}\v_{i_5}]
-[\v_{i_1}\v_{i_3}][\v_{i_2}\v_{i_4}\v_{i_5}]\\

+[\v_{i_1}\v_{i_4}][\v_{i_2}\v_{i_3}\v_{i_5}]
-[\v_{i_1}\v_{i_5}][\v_{i_2}\v_{i_3}\v_{i_4}]
&=& 0.
\ea
\ee

\item[$\rm DB:$] For any six symbols $\v_{i_1}, \ldots, \v_{i_6}$,
\be\hskip -.75cm
[\v_{i_1}\v_{i_2}\v_{i_3}][\v_{i_4}\v_{i_5}\v_{i_6}]
=-\left|\ba{lll}
{[}\v_{i_1}\v_{i_4}] & [\v_{i_1}\v_{i_5}] & [\v_{i_1}\v_{i_6}]\\
{[}\v_{i_2}\v_{i_4}] & [\v_{i_2}\v_{i_5}] & [\v_{i_2}\v_{i_6}]\\
{[}\v_{i_3}\v_{i_4}] & [\v_{i_3}\v_{i_5}] & [\v_{i_3}\v_{i_6}]
\ea
\right|.
\ee
\ed

The {\it inner-product bracket algebra} is the
commutative ring generated by the above two kinds of symbols, satisfying the symmetry
requirements and the dimension-three constraints.

To include brackets of longer length, the concept quaternionic bracket algebra or
Clifford bracket algebra needs to be introduced. As explained before, there is no need to 
distinguish between the two concepts in the setting of 3D orthogonal geometry, so we simply call it
{\it bracket algebra}. To distinguish from the concept of the same name arising from Grassmann-Cayley algebra
\cite{white}, we call that in \cite{white} {\it classical bracket algebra}. 

Formally, besides the above 2-tuples and 3-tuples, a hierarchy of infinitely many new symbols can be defined:
for any length $l>3$, there are all $l$-tuples of symbols 
selected repetitively from ${\cal M}$, with the requirement that the first and the last equalities in 
Caianiello expansion (\ref{eq4}) are satisfied; such an $l$-tuple is denoted by 
$[\v_{j_1}\v_{j_1}\cdots \v_{j_l}]$. By setting $[1]=1$ (0-tuple) and $[\v_i]=0$ (1-tuple) for all $i$,
we get a full hierarchy of new symbols marked by brackets, with arbitrary length $l\geq 0$.
The new symbols together with their specific requirements,
form a a commutative ring called the
{\it bracket algebra} over 3D inner-product space. This is the the bottom-up approach
to defining bracket algebra.
The concept quaternionic product or Clifford product is not needed.

\section{Vector-variable polynomials \\
and bracket polynomials}
\setcounter{equation}{0}
\vskip .2cm

In this paper, we use (1) bold-faced digital numbers
and bold-faced lower-case letters to denote vector variables, {\it e.g.}, $\v, \1$; (2) 
bold-faced upper-case letters to denote monic vector-variable monomials, {\it e.g.}, $\A, \X$; 
(3) Roman-styled
lower-case letters to denote polynomials, {\it e.g.}, 
$f,g$; (4) Greek letters
to denote $\mathbb K$-coefficients, {\it e.g.}, $\lambda, \mu$.

Although the background is real orthogonal geometry, the algebraic manipulations under investigation
are independent of the real field. In fact, only the following coefficients occur in computing:
$\pm 2^k$ for $k\in \mathbb Z$. We set the base field $\mathbb K$ to be of characteristic $\neq 2$.

Now start from quaternions. Let $\v_{1}, \v_{2}, \ldots, \v_{n}$ be symbols. 
What properties determine that the multilinear associative product among the symbols
is the quaternionic one, and that these symbols represent vectors
of a 3D real inner-product space with metric diag$(-1,-1,-1)$? \cite{li13} gives a rather simple answer.

The inner product of two vectors $\v_i, \v_j$
is a scalar, so it commutes with a third vector $\v_k$:
$[\v_i\v_j]\v_k=\v_k[\v_i\v_j]$.
For three vectors $\v_i, \v_j, \v_k$, since $[\v_i\v_j\v_k]$ is a scalar, for a fourth vector $\v_l$,
the commutativity
$[\v_i\v_j\v_k]\v_l=\v_l[\v_i\v_j\v_k]$ holds. The two commutativities are all that characterize
the equality properties of the quaternionic product, besides the  
multilinearity and associativity.

\bt ${\rm \cite{li13}}$
Let $\v_1, \v_2, \ldots, \v_n$ be $n>2$ symbols. Define
the product among them, denoted by juxtaposition of elements, as the 
$\mathbb K$-tensor product modulo the 
two-sided ideal generated by the following tensors:
\be\hskip -.15cm
\ba{ll}
{\rm V2}:   &\hskip .1cm \v_i\otimes \v_i\otimes \v_j- \v_j\otimes \v_i\otimes \v_i; \\
{\rm V3}:   &\hskip .1cm (\v_i\otimes \v_j+\v_j\otimes \v_i)\otimes \v_k-\v_k\otimes (\v_i\otimes \v_j+\v_j\otimes \v_i); \\
{\rm V4}:   &\hskip .1cm (\v_i\otimes \v_j\otimes \v_k-\v_k\otimes \v_j\otimes \v_i)\otimes \v_l\\
& \hskip .8cm
-\v_l\otimes (\v_i\otimes \v_j\otimes \v_k-\v_k\otimes \v_j\otimes \v_i),
\ea
\label{eq0}
\ee
for any $i\neq j\neq k\neq l$ in $1,2,\ldots, n$. 
Denote by ${\cal Q}$ or ${\cal Q}[[\v_1, \ldots, \v_n]]$ the $\mathbb K$-algebra defined by the above product 
and generated by the $\v_i$. Denote by
${\cal I}[[\v_1, \ldots, \v_n]]$ the above ideal, and call it the {\it syzygy ideal} of
${\cal Q}$. 

Denote 
\be\ba{lll}
{\mathbb K}_2 &:=& {\mathbb K}(\{\v_i\otimes \v_i,\ \v_i\otimes \v_j+\v_j\otimes \v_i, \,|i\neq j\}),\\
{\mathbb K}_3 &:=& {\mathbb K}_2(\{ 
\v_i\otimes \v_j\otimes \v_k-\v_k\otimes \v_j\otimes \v_i\,|\,i\neq j\neq k\}).
\ea
\label{q:field}
\ee
 
(1) Let $\iota:=[\v_1\v_2\v_3]$. Then $\iota\neq 0$, and ${\mathbb K}_3= {\mathbb K}_2(\iota)$.

(2)  
The $\v_i$ and $\iota\v_i\times \v_j$ of the $\mathbb K$-algebra ${\cal Q}$
span a 3D ${\mathbb K}_2$-space ${\cal V}^3$. 
For any $k\geq 1$, $A(\v_{j_1}\cdots \v_{j_{2k+1}})$ 
and $\iota A(\v_{j_1}\cdots \v_{j_{2k}})$
are both in ${\cal V}^3$.

(3) The defined product is the Clifford product of the
${\mathbb K}_2$-Clifford algebra over ${\cal V}^3$.

(4) If ${\mathbb K}={\mathbb K}_3=\mathbb R$, and the inner product of real space
${\cal V}^3$ induced from (\ref{inner}) is definite, then
the defined product is the quaternionic product.

(5) Let $\V_k=\v_{j_1}\v_{j_2}\cdots \v_{j_k}$. Then
the following identities hold modulo ${\cal I}[[\v_1, \v_2, \ldots, \v_n]]$: 
\be
\v_i[\V_k]=[\V_k]\v_i,\ \ \,
[\v_i\V_k]=[\V_k\v_i]. \label{general:shift}
\ee
\et

The requirements in (4) distinguishing the quaternionic product from the Clifford product
cannot be represented by equalities. So for symbolic manipulations of equalities,
the quaternionic product and the Clifford product cannot be distinguished.
The $\mathbb K$-algebra $\cal Q$ is called the {\it 3D vector-variable polynomial ring} 
generated by vector variables $\v_i$, and the product in it is called the 
{\it vector-variable product}. It is neither the quaternionic product nor the Clifford product,
but a more basic one.

All the terminologies introduced earlier on quaternions and Clifford algebra are valid for
$\cal Q$. Besides, there are some additional terminologies for $\cal Q$.
Let $\v_1\prec \v_2\prec \ldots\prec \v_n$ be vector variables. 
A {\it monic monomial} of vector variables
refers to the product of a repetitive permutation of some of the 
vector variables. 
For a monic monomial 
$\v_{i_1}\v_{i_2}\cdots \v_{i_k}$, the {\it leading variable} refers to $\v_{i_1}$, and the 
{\it trailing variable}
refers to $\v_{i_k}$. The monomial is said to be {\it non-descending} if 
$\v_{i_1}\preceq \v_{i_2}\preceq \ldots\preceq \v_{i_k}$,
and is said to be
{\it ascending} if $\v_{i_1}\prec \v_{i_2}\prec \ldots\prec \v_{i_k}$. 
The {\it degree}, or {\it length}, of the monomial is $k$.
The {\it lexicographic ordering} among monomials is always assumed.

A {\it polynomial} of vector variables is a $\mathbb K$-linear combination of 
monic monomials. The {\it leading term} of a polynomial $f$ is the term of highest order, 
denoted by $\lt(f)$.
The {\it degree} of a polynomial is that of its leading term. 
The leading terms of all elements in a subset $\cal S$ of polynomials
are denoted by $\lt({\cal S})$. When specifying the field ${\mathbb K}_2$ or 
${\mathbb K}_3$, we can get the corresponding concepts {\it quaternionic polynomial}
and {\it Clifford polynomial}.

Fix a multiset of vector variables $\cal M$ composed of $m\geq 3$ 
symbols $\v_1, \v_2, \ldots, \v_m$. Let $n$ be the number of different elements in $\cal M$,
where $3\leq n\leq m$. 
In the $\mathbb K$-tensor algebra $\bigotimes(\v_1, \v_2, \ldots, \v_n)$ generated by the $n$ symbols
taken as vectors,
a {\it tensor monomial} is up to coefficient the tensor product of finitely many such vectors.
The {\it $\mathbb K$-tensor algebra} over $\cal M$, denoted by $\bigotimes[{\cal M}]$, is the 
$\mathbb K$-subspace
of $\bigotimes(\v_1, \v_2, \ldots, \v_n)$ spanned by tensor monomials whose vector variables
by counting multiplicity are in $\cal M$, equipped with the tensor product that is undefined if the 
result is no longer in $\bigotimes[{\cal M}]$.

When the product among the elements in $\cal M$ is the vector-variable product, we 
have the corresponding concept of
{\it 3D vector-variable polynomial ring over multiset} $\cal M$, denoted by ${\cal Q}[{\cal M}]$. 
Each element in ${\cal Q}[{\cal M}]$
is a polynomial whose multiset of vector variables in each term is a 
submultiset of $\cal M$.
The syzygy ideal ${\cal I}[{\cal M}]$ of ${\cal Q}[{\cal M}]$ is still defined by
(\ref{eq0}).

The concepts of {\it Gr\"obner base} and {\it normal form} are defined in 
${\cal Q}[{\cal M}]$
just as in $\bigotimes(\v_1, \v_2, \ldots, \v_n)$ \cite{mora}.
For two monomials $h_1, h_2$ in vector variables, 
$h_1$ is said to be {\it reduced} with respect to $h_2$, if 
$h_2$ is not a {\it factor} of $h_1$, or $h_1$ is not a {\it multiplier} of $h_2$,
{\it i.e.}, there do not exist monomials $l,r$, including elements of $\mathbb K$, such that 
$h_1=lh_2r$. For two polynomials $f$ and $g$, $f$ is said to be {\it reduced} with respect to $g$,
if the leading term of $f$ is reduced with respect to that of $g$. 
The term ``{\it non-reduced}" means the opposite.

Let $\{f_1, f_2, \ldots, f_k\}$ be a set of vector-variable polynomials. 
Another set of vector-variable polynomials
$\{g_1, g_2, \ldots$, $g_l\}$ is said to be a {\it reduced Gr\"obner base} of the ideal 
${\cal I}:=\langle f_1, f_2, \ldots, f_k \rangle$ generated by the $f_i$ in $\cal Q[{\cal M}]$,
if 
(1) $\langle g_1, \ldots, g_l \rangle={\cal I}$, (2) the leading term of any
element in $\cal I$ is a multiplier of the leading term of some $g_i$, (3) the $g_i$ are pairwise reduced
with respect to each other.

The {\it reduction} of a polynomial $f$ with respect to a reduced Gr\"obner base
$g_1, g_2, \ldots, g_l$ is the repetitive procedure of dividing the highest-ordered 
non-reduced term $L$ of $f$ by some 
$g_i$ whose leading term is a factor of $L$, then updating $f$ by replacing
$L$ with its remainder, until all terms of $f$ are reduced. 
The result is called the {\it normal form} of $f$ with respect to
the Gr\"obner base. Two polynomials are equal if and only if they have 
identical normal forms.

In \cite{li13}, two theorems are established for the 
Gr\"{o}bner base and normal forms of 3D vector-variable polynomials, one for the 
multilinear case where each element in multiset $\cal M$ has multiplicity 1, 
the other for the general case ${\cal Q}[[\v_1, \ldots, \v_n]]$. 

\bt  \label{main:1} ${\rm \cite{li13}}$
Let 
${\cal I}[\v_1, \ldots, \v_n]$ be the syzygy ideal of the multilinear polynomial ring 
${\cal Q}[\v_1, \ldots, \v_n]$ in $n$ different vector variables $\v_1\prec \v_2\prec \ldots \prec \v_n$. 

(1) [Gr\"obner base] The following are a reduced Gr\"obner base of ${\cal I}[\v_1, \ldots, \v_n]$:
for all $1\leq i_1<i_2<\ldots<i_j\leq n$,
\bu
\item[$\rm G3$:] 
${[}\v_{i_3}\v_{i_2}\v_{i_1}]-[\v_{i_1}\v_{i_3}\v_{i_2}]$, and 
${[}\v_{i_3}\v_{i_1}\v_{i_2}]-[\v_{i_2}\v_{i_3}\v_{i_1}]$;

\item[${\rm G}j$:] 
${[}\v_{i_3}\v_{i_2}\v_{i_4}\v_{i_5}\cdots \v_{i_j}\v_{i_1}]
-{[}\v_{i_2}\v_{i_4}\v_{i_5}\cdots \v_{i_j}\v_{i_1}\v_{i_3}]$, for all $j>3$.
\eu

(2) [Normal form] In a normal form, every term is
up to coefficient 
of the form $\V_{Y_1}\v_{z_1}\V_{Y_2}\v_{z_2}\cdots \V_{Y_k}\v_{z_k}$ or 
$\V_{Y_1}\v_{z_1}$ $\cdots \V_{Y_{k}}\v_{z_{k}}\V_{Y_{k+1}}$, where \\
(i) $k\geq 0$, \\
(ii) $\v_{z_1}\v_{z_2}\cdots \v_{z_k}$ 
is ascending, \\
(iii) every 
$\V_{Y_i}$ is an ascending monomial of length $\geq 1$, \\
(iv) 
$\V_{Y_1}\V_{Y_2}\cdots \V_{Y_k}$ (or $\V_{Y_1}\V_{Y_2}\cdots \V_{Y_{k+1}}$ if $\V_{Y_{k+1}}$ occurs) is ascending, \\
(v) for every $i\leq k$, if $\v_{t_i}$ is the trailing variable of 
monomial $\V_{Y_i}$, then $\v_{z_i}\prec \v_{t_i}$.
\et

\bt \label{main:2} ${\rm \cite{li13}}$
Let ${\cal I}[[\v_1, \ldots, \v_n]]$ be the syzygy ideal of the
polynomial ring ${\cal Q}[[\v_1, \ldots, \v_n]]$
in $n$ different vector variables $\v_1\prec \v_2\prec \ldots \prec \v_n$. 

(1) [Gr\"obner base] The following are a reduced Gr\"obner base of ${\cal I}[[\v_1, \ldots, \v_n]]$:
\bu
\item[$\rm G3$, \ ] \hskip -.3cm${\rm G}j$:\ \, for all $3<j<\infty$, and all
$1\leq i_1<i_2<i_3< i_4\leq \ldots \leq i_{j-1}<i_j\leq n$;

\item[$\rm EG2$:] for all $i_1<i_2$,
\[\ba{cc}
\v_{i_2}\v_{i_2}\v_{i_1}-\v_{i_1}\v_{i_2}\v_{i_2}, &
\v_{i_2}\v_{i_1}\v_{i_1}-\v_{i_1}\v_{i_1}\v_{i_2}; 
\ea
\]

\item[${\rm EG}j$:]
$
{[}\v_{i_3}\v_{i_2}\v_{i_3}\v_{i_4}\cdots \v_{i_j}\v_{i_1}]
-{[}\v_{i_2}\v_{i_3}\v_{i_4}\cdots \v_{i_j}\v_{i_1}\v_{i_3}]
$,\
for all $2< j<\infty$, and all
$1\leq i_1<i_2<i_3\leq i_4\leq \ldots \leq i_{j-1}<i_j\leq n$.
\eu

(2) [Normal form] In a normal form, every term is up to coefficient
of the form $\V_{Y_1}\v_{h_1}\v_{z_1}\V_{Y_2}\v_{h_2}\v_{z_2}\cdots \V_{Y_k}\v_{h_k}\v_{z_k}$ or
$\V_{Y_1}\v_{h_1}\v_{z_1}\cdots \V_{Y_k}\v_{h_k}\v_{z_k}\V_{Y_{k+1}}$,
where \\
(i) $k\geq 0$, \\
(ii) $\v_{z_1}\v_{z_2}\cdots \v_{z_k}$ is non-descending, \\
(iii) $\v_{h_1}\v_{h_2}\cdots \v_{h_k}$ 
is non-descending, \\
(iv) every 
$\V_{Y_i}$ is a non-descending monomial of length $\geq 0$, \\
(v) 
$\V_{Y_1}\v_{h_1}\V_{Y_2}\v_{h_2}\cdots \V_{Y_k}\v_{h_k}$ 
(or $\V_{Y_1}\v_{h_1}\cdots \V_{Y_k}\v_{h_k}\V_{Y_{k+1}}$
if $\V_{Y_{k+1}}$ occurs) is non-descending, 
\\
(vi) for every $i\leq k$, $\v_{h_i}\succ \v_{z_i}$,
\\
(vii) for every $i\leq k$, 
if the length of $\V_{Y_i}$ is nonzero, let
$\v_{t_i}$ be the trailing variable of $\V_{Y_i}$, then
$\v_{t_i}\prec \v_{h_i}$.
\et

For a general multiset $\cal M$ in which the different vector variables are 
$\v_1, \v_2, \ldots, \v_n$, the Gr\"obner base of the syzygy ideal 
${\cal I}[{\cal M}]$ is the restriction of
the Gr\"obner base of ${\cal I}[[\v_1, \ldots, \v_n]]$ to ${\cal Q}[{\cal M}]$,
denoted by ${\cal G}[{\cal M}]$. In ${\cal Q}[{\cal M}]$,
a polynomial is said to be {\it $\cal I$-normal} if its leading term is 
reduced with respect to the Gr\"obner base. The procedure of deriving the normal form of
a polynomial is called {\it $\cal I$-reduction}.

Now that any vector-variable polynomial has a normal form by $\cal I$-reduction, 
so does a bracket polynomial when every
bracket is expanded into two terms by definition. The result is complicated. 

Consider the following simple example: for a single bracket
\[
2[\v_1\v_2\cdots \v_m]=\v_1\v_2\cdots \v_m+(-1)^m\v_m\cdots \v_2\v_1,
\]
the $\cal I$-reduction goes as follows: 
for $0\leq j<m$, if we define $\V_{m-j}=\v_{j+1}\v_{j+2}\cdots \v_m$, then when $m\geq 3$, 
\be\ba{ll}
& \V_m+(-1)^m\V_m^\dagger \\
\stackrel{{\cal I}}{=} &  \v_1\V_{m-1}+\V_{m-1}\v_1-\v_1(\V_{m-1}+(-1)^{m-1}\V_{m-1}^\dagger)\\
\stackrel{{\cal I}}{=} &  \v_1\V_{m-1}+\V_{m-1}\v_1-\v_1(\v_2\V_{m-2}+\V_{m-2}\v_2)\\

& \hfill
+\v_1\v_2(\V_{m-2}+(-1)^{m-2}\V_{m-2}^\dagger)\\

=& \v_2\V_{m-2}\v_1-\v_1\V_{m-2}\v_2\\

& \hfill
+\v_1\v_2(\V_{m-2}+(-1)^{m-2}\V_{m-2}^\dagger).
\ea
\ee
From this recursive formula, we get for $m\geq 1$, 
\be\ba{lcl}
2[\v_1\v_2\cdots \v_m] &\hskip -.1cm\stackrel{{\cal I}}{=}& \hskip -.1cm
\ds \frac{1+(-1)^m}{2}\v_1\v_2\cdots \v_m\\

&& \ds +\sum_{i=1}^{m-1} (-1)^{i+1} (\v_1\cdots \check{\v_i}\cdots \v_m)\v_i.\Bigstrut
\ea
\ee

So the normal form of the simplest bracket $[\v_1\v_2\cdots \v_m]$ is composed of up to 
$m$ terms. What is worse is that only when the terms are summed up can they represent a single
algebraic invariant (the bracket), while missing a single term destroys the invariance of the whole 
expression. 

From the appearance, the bracket symbol only hides half of a binomial. There are much more behind this
appearance. By definition, for a sequence $\A$ of $a>1$ vector variables, its bracket 
is $[\A]=2^{-1}\A+(-1)^a2^{-1}\A^\dagger$. Monomial $\A$ is called the {\it representative} 
of bracket $[\A]$. Later on, when we write $[\A]$, we always assume that monomial $\A$ is 
the representative of the bracket. 

The definition of a bracket endows the symbol with the reversion symmetry 
(or equivalently, the conjugate symmetry)
up to sign:
$[\v_{i_1}\ldots \v_{i_a}]=(-1)^{a}[\v_{i_a}\cdots \v_{i_1}]$. By 
(\ref{general:shift}), the bracket symbol also has shift symmetry.
So up to sign a bracket of $a$ vector variables has the symmetry group
$D_{2a}$ (dihedral group). 

The above analysis is only for a single bracket. For bracket polynomials, there are a lot of
polynomial identities, or {\it syzygies}, among them. These complexities justify the separation of
bracket algebra from vector-variable polynomial ring in symbolic manipulations of algebraic invariants.
Finding the Gr\"obner base of the syzygies 
and then characterizing the normal forms of bracket polynomials are the main goal of this paper.

The following are some terminology on brackets.
The {\it representative} of a bracket polynomial is the 
the vector-variable polynomial
whose terms are each the product of the coefficient with
the representatives of the bracket factors in the same term. 
The representative of a bracket polynomial 
is allowed to contain brackets. For example, $[\v_1\v_2[\v_3\v_4]]$ is
taken as bracket binomial $2^{-1}[\v_1\v_2(\v_3\v_4+\v_4\v_3)]$; its representative is the content
$\v_1\v_2[\v_3\v_4]$ within the outer bracket. 

The {\it lexicographic ordering} of bracket polynomials
is that of their representatives. 
The {\it leading variable} of a bracket refers to that of its representative.
The {\it leading term} of a bracket polynomial is always under the lexicographic ordering.
For example, if $\v_1\prec \v_2$, then $[\v_1\v_2]\prec [\v_2\v_1]$, and
$[\v_1\v_2][\v_2\v_1]\prec [\v_2\v_1][\v_1\v_2]$.

The {\it leader} (or {\it expanded leading term}) 
of a bracket polynomial,
refers to the leading term of the bracket polynomial when taken as a vector-variable one,
{\it i.e.},
the corresponding vector-variable polynomial obtained from expanding each bracket
into two terms by definition. For example, the leader of bracket $[\A]$
refers to the one of higher order between 
$2^{-1}\A$ and $(-1)^a2^{-1}\A^\dagger$. 

Among all the bracket polynomials that are equal to the same bracket polynomial, there are two 
that have strong features: the first is the one whose representative is the lowest,
the second is the one whose leader is the lowest.
The second is unique but the first is not. To make the first unique we 
introduce the following concept.

In ${\cal Q}[{\cal M}]$, where the number of elements in multiset $\cal M$ is $m$, 
a {\it uni-bracket monomial} refers to a single bracket of length $m$.
A {\it uni-bracket polynomial} is a $\mathbb K$-linear combination of uni-bracket monomials.
All uni-bracket polynomials form a $\mathbb K$-linear space, denoted by $[{\cal Q}][{\cal M}]$.
The $\mathbb K$-linear space of the representatives of elements in
$[{\cal Q}][{\cal M}]$ is just the $\mathbb K$-linear space of degree-$m$ vector-variable polynomials,
denoted by ${\cal Q}_m[{\cal M}]$. Obviously, $[{\cal Q}][{\cal M}]$ is a linear subspace of 
${\cal Q}_m[{\cal M}]$.

When taken as a vector-variable polynomial, a uni-bracket is a binomial. In appearance, a uni-bracket
is a monomial. To distinguish between the two understandings, we need a device to get rid of the bracket
symbol and extract the representative of the uni-bracket. 
This can be done by taking $[{\cal Q}][{\cal M}]$ as the quotient of ${\cal Q}_m[{\cal M}]$
modulo the ideal 
\be
{\cal J}[{\cal M}]:={\cal I}[{\cal M}]+[{\cal I}][{\cal M}],
\ee
where $[{\cal I}][{\cal M}]$ is
composed of the vector parts of degree-$m$ 
polynomials, {\it i.e.}, the $\mathbb K$-linear span of elements of the form
\be
{\rm R}:\ \ \v_{i_1}\v_{i_2}\ldots \v_{i_m}-(-1)^{m}\v_{i_m}\cdots \v_{i_2}\v_{i_1},
\label{general:R} 
\ee
for all permutations of the $m$ elements in $\cal M$. 

The modulo-$[{\cal I}][{\cal M}]$ operation identifies a uni-bracket with its representative, or equivalently,
identifies any degree-$m$ vector-variable polynomial with the uni-bracket polynomial it serves as the representative.
This operation retains the bracket symbols of all the brackets of length $<m$, while removing the bracket
symbols from all brackets of length $m$.

\section{Gr\"obner base and normal form for multilinear uni-bracket \\
polynomials}
\setcounter{equation}{0}
\vskip .2cm

From this section on, we use bold-faced digital numbers to denote vector variables, and use
bold-faced capital letters to denote monic monomials of vector variables.

In this section, the multiset $\cal M$ is composed of $m\geq 3$ 
different vector variables, and the 
modulo-$[{\cal I}][{\cal M}]$ operation is always assumed. Then $[{\cal Q}][{\cal M}]$ and ${\cal Q}_m[{\cal M}]$
are identical, and a uni-bracket no longer has the outer bracket symbol. Ideal $[{\cal I}][{\cal M}]$
is called the {\it uni-bracket removal ideal} in ${\cal Q}[{\cal M}]$, and ideal 
${\cal J}[{\cal M}]$ is called the {\it syzygy ideal} of $[{\cal Q}][{\cal M}]$
in $\bigotimes[{\cal M}]$. 
Below we compute the Gr\"obner base of ${\cal J}[{\cal M}]$
and characterize the normal forms of uni-bracket polynomials.

The following are elements of ${\cal J}[{\cal M}]$: for all monomials $\A$ to $\F$ such that
$\A\1\B, \1\C, \1\D\2, \1\E\2\F$ are of length $m$ and $\E$ has length $e>0$,
\be\ba{ll}
{\rm S}_1: & \A\1\B-\1\B\A,\\

{\rm S}_1^N: &  \hbox{the ${\rm S}_1$ in which $\A\1\B$ is $\cal I$-normal},\\

{\rm R}_1: & 2(\1[\C])=\1\C-(-1)^{m}\1\C^\dagger, \\

{\rm R}_{12}: & 2(\1[\D\2])=\1\D\2-(-1)^{m}\1\2\D^\dagger, \\

{\rm R}_{12}[*]: & 2(\1[\E\2[\F]])=\1\E\2[\F]-(-1)^{e}\1\2\E^\dagger[\F].
\ea 
\ee

Since the reduced Gr\"obner base of ${\cal I}[{\cal M}]$ is ${\cal G}{[}{\cal M}{]}$, we only need to consider the
elements of type $\rm R$ in ${\cal J}[{\cal M}]$, as they span $[{\cal I}][{\cal M}]$. By
\be\ba{ll}
& \A\1\B-(-1)^{m}\B^\dagger\1\A^\dagger \\
=& (\A\1\B-\1\B\A)+(\1\B\A
-(-1)^{m}\1\A^\dagger\B^\dagger)\\

&\hfill
-(-1)^m(\B^\dagger\1\A^\dagger-\1\A^\dagger\B^\dagger),
\ea
\label{estab:r}
\ee
we get

\bl
${\rm R}$ is a subset of the ideal $\langle {\rm S}_1, {\rm R}_1\rangle$. 
For any type-$\rm R$ element
$f$ but not of type ${\rm R}_{1}$, $\lt(f)\in \lt({\rm S}_1)$.
\el

\bl\label{lem:s}
${\rm S}_1$ is a subset of $\langle{\rm S}_1^N, {\rm R}_1\rangle+{\cal I}[{\cal M}]$. 
For any type-${\rm S}_1$ element
$f$ but not of type ${\rm S}_1^N$, $\lt(f)\in \lt({\cal I}[{\cal M}])$.
\el

\bproof 
Consider a general type-${\rm S}_1$ element $f=\A\1\B-\1\B\A$, where $\A$ is not empty. 
If $\A\1\B$ is $\cal I$-normal, then $f\in {\rm S}_1^N$.
If $\cal I$-reductions are
carried out to $\A, \B$, say $\A=\A^N+{\cal I}$ and
$\B=\B^N+{\cal I}$, where $\A^N, \B^N$ are both $\cal I$-normal,
then $f=\A^N\1\B^N-\1\B^N\A^N+{\cal I}$, and $\A^N\1\B^N$ is the new leading term. 
So by $\cal I$-reductions
we can assume that both $\A$ and $\B$ are $\cal I$-normal. 

Assume that $\A\1\B$ is not $\cal I$-normal. Further
assume that any type-${\rm S}_1$ element $g\prec f$ is in 
$\langle{\rm S}_1^N, {\rm R}_1\rangle+{\cal I}[{\cal M}]$.
We prove the conclusion for $f$ by reduction on the order. 
There are three possibilities to
apply Gr\"obner base elements of ${\cal I}[{\cal M}]$ to make reduction to $\A\1\B$:
(i) G3 at the end of $\A\1$, (ii) G$i$ for $i>3$ at the end of $\A\1$, (iii) G3 on $\1$ and two variables from $\A,\B$
respectively.

Case (i). Let $\A\1\B=\C\u\v\1\B$, where $\u\succ\v\succ \1$. Then
\[\ba{lcl}
f &\stackrel{{\rm G3}}{=}& 
-\C\v\u\1\B+\C\1(\u\v+\v\u)\B-\1\B\C\u\v\\

&\stackrel{{\rm induction}}{=}& 
-\1\B\C(\v\u+\u\v)+\1(\u\v+\v\u)\B\C
\\

&\stackrel{{\cal I}}{=}& 
0.
\ea
\]

\def\localgap{\hskip -.3cm}

Case (ii). Let $\A\1\B=\C\u\v\D\1\B$, where $1\prec \v\prec\u\prec$ all variables of $\D$. 
Let the length of $\D$ be $d>0$. Let the lengths of $\B, \C$ be $b,c$ respectively. Then $b+c=m-d-3$.
\[\ba{lcl}
f &\localgap\stackrel{{\rm G}i}{=}& \localgap
\C(\v\D\1+(-1)^d \1\D^\dagger\v)\u\B-(-1)^d \C\u\1\D^\dagger\v\B\\

&& \hfill
-\1\B\C\u\v\D\\

&\localgap\stackrel{{\rm induction}}{=}& \localgap
\1\{\underline{(\u\B\C-\B\C\u)\v\D}+(-1)^d \D^\dagger\v(\u\B\C-\B\C\u)\}
\\

&\localgap\stackrel{{\rm R}_1}{=}& \localgap
(-1)^d \1\D^\dagger\v\{
(-1)^{m-d}(\C^\dagger\B^\dagger\u-\u\C^\dagger\B^\dagger)\\

&& \hfill
+(\u\B\C-\B\C\u)
\}
\\

&\localgap =& \localgap
(-1)^d \1\D^\dagger\v\{
-((-1)^{b+c}\C^\dagger\B^\dagger+\B\C)\u\\

&& \hfill
+\u((-1)^{b+c}\C^\dagger\B^\dagger+\B\C)\}
\\

&\localgap\stackrel{{\cal I}}{=}& \localgap
0.
\ea
\]

Case (iii). Let
$\A\1\B=\C\u\1\v\D$, where $\u\succ \v \succ \1$. Let the lengths of $\C, \D$ be $c,d$ respectively.
Then $c+d=m-3$.
\[\ba{lcl}
f &\stackrel{{\rm G3}}{=}& 
-\C\1\u\v\D+\C\v(\u\1+\1\u)\D-\1\v\D\C\u
\\

&\stackrel{{\rm induction}}{=}& 
\1(-\underline{\u\v\D\C}+\D\C\v\u+\underline{\u\D\C\v}-\v\D\C\u)
\\

&\stackrel{{\rm R}_1}{=}& 
\1\{-((-1)^m\C^\dagger\D^\dagger-\D\C)\v\u\\

&& \hfill
+\v((-1)^m\C^\dagger\D^\dagger-\D\C)\u\}
\\

&\stackrel{{\cal I}}{=}& 
0.
\ea
\]
\endproof

\bl
${\rm R}_1$ is a subset of the ideal $\langle {\rm R}_{12}\rangle+{\cal I}[{\cal M}]$. 
For any type-${\rm R}_1$ element
$f$ but not of type ${\rm R}_{12}$, 
$\lt(f)\in \lt({\cal I}[{\cal M}])$.
\el

\bproof
Let $f=2(\1[\E\2\F])$ be a general element of type ${\rm R}_1$. Then
$f\stackrel{{\cal I}}{=}2(\1[\F\E\2])$, and the latter is in ${\rm R}_{12}$.
\endproof

\bt 
\label{main:bracket}
Let ${\cal M}=\{\1,\2,\ldots, \m\}$ be $m>2$ different symbols, where
$\1\prec\2\prec \ldots\prec \m$, and let 
${\cal Q}[{\cal M}]$ be the vector-variable polynomial ring over $\cal M$.
Let 
$[{\cal Q}][{\cal M}]$ be the space of uni-bracket polynomials, and let 
${\cal J}[{\cal M}]$
be its syzygy ideal in $\bigotimes[{\cal M}]$.

(1) [Gr\"obner base] The following are a reduced Gr\"obner base of ${\cal J}[{\cal M}]$: 

\bu
\item[${\cal G}{[}{\cal M}{]}$:] $\rm G$i for all $3\leq i\leq m$;

\item[${\rm S}_1^N$:] $\A\1\B-\1\B\A$, where $\A$ is an ascending sequence of length $a>0$, 
$\B$ is of length $m-a-1\geq 0$, and $\A\1\B$ is $\cal I$-normal;

\item[${\rm R}_{12}^N\hbox{$\rm [$}j\hbox{$\rm ]$}$:] for all $1\leq j\leq \frac{m-1}{2}$, \
$\1[\A\2][\Y_2\z_2][\Y_3\z_3]\cdots[\Y_j\z_j]$,
where (i) $\A$ is an ascending sequence of length $a>0$;
\\
(ii) when $j=1$, then ${\rm R}_{12}^N[1]=\1[\A\2]$;
\\
(iii) each $\Y_i$ is a non-empty ascending sequence, and each $\z_i$ is a variable such that
$\z_i\Y_i$ is ascending;
\\
(iv) $\1\A\2\Y_2\z_2\cdots\Y_j\z_j$ is $\cal I$-normal and length-$m$.
\eu

(2) [Normal form] In a normal form, every term is $\cal I$-normal, and is up to coefficient
of one of the following forms: \\
(I) $\1\2\C$, where $\C$ is $\cal I$-normal;\\
(II) $\1\A\2\Y_2\z_2\Y_3\z_3\cdots \Y_{k-1}\z_{k-1}\Y_{k}$, where $k\geq 2$, 
$\A$ and the $\Y_i$ are each a non-empty ascending sequence, and each $\z_i\prec \t_i$, the latter being
 the trailing
variable of $\Y_i$;
\\
(III)
$\1\A\2\Y_2\z_2\Y_3\z_3\cdots \Y_{k}\z_{k}$, where $k\geq 2$, 
$\A$ and the $\Y_i, \z_i$ are as in (II),
and for some $2\leq i\leq k$, 
if $\l_{i}$ is the leading variable of $\Y_{i}$, then
$\l_{i}\prec \z_{i}$.

\et

\bproof There are several steps.

{\it Step 1}. We need to prove that
${\rm R}_{12}$ is a subset of the ideal 
$\langle {\rm R}_{12}^N\rangle+{\cal I}[{\cal M}]$. Once this is done, then 
since the leader of 
any element of ${\rm R}_{12}^N$ is ${\cal I}$-normal and 
cannot be cancelled by the leader of 
any other element of ${\rm R}_{12}^N$, the ${\rm G}i$ and 
${\rm R}_{12}^N[j]$ form a reduced Gr\"obner base of 
$\langle {\rm R}_{12}\rangle+{\cal I}[{\cal M}]$. By this and the previous three lemmas,
we get conclusion (1) of the theorem. 

Once conclusion (1) holds, then any $\cal I$-normal 
monomial of length $m$  with leading variable $\1$ is the representative of a uni-bracket
in normal form if and only if it is not the leader
of an ${\rm R}_{12}^N$-typed element. Conclusion (2) follows.

{\it Step 2}. The idea of proving the statement in Step 1 is to use
${\cal G}{[}{\cal M}{]}$ to decrease the order of the leader of every element of type 
${\rm R}_{12}$ or ${\rm R}_{12}[*]$, at the same time keep the reduction result to be within the 
$\mathbb K$-linear space spanned by elements of type
${\rm R}_{12}$ or ${\rm R}_{12}[*]$. Then ultimately all the leaders of these elements
become $\cal I$-normal.

We start with the $\cal I$-reduction on the leading term of 
a general ${\rm R}_{12}$-typed element $f=\1\A\2-(-1)^m\1\2\A^\dagger$, where
the length of $\A$ is $m-2$. If by 
$\cal I$-reduction, $\A=\A^N+{\cal I}$, then $f=\1\A^N\2-(-1)^m\1\2{\A^N}^\dagger+\cal I$.
So we can assume that $\A$ is $\cal I$-normal.

If $\A\2$ is $\cal I$-normal, then $f$ is just ${\rm R}_{12}^N[1]$. When $\A\2$ is not $\cal I$-normal,
if $\A\2$ is non-reduced with respect to G3, let $\A\2=\B\u\v\2$, where 
$\u\succ \v\succ \2$,
then
\be
\1\B\u\v\2=-\1\B\v\u\2+2\{\1\B\2[\u\v]\}.
\label{g3:proof}
\ee
The result consists of the leading terms of one ${\rm R}_{12}$-typed element
and one ${\rm R}_{12}[*]$-typed element. The leaders of both terms are lower than $f$.

\def\localgap{\hskip -.1cm}

If $\A\2$ is non-reduced with respect to G$i$ for some $i>3$, let $\A\2=\B\u\v\C\2$, where 
$\u\succ \v\succ \2$, and $\u\C$ is ascending, and the length of $\C$ is $c>0$. Then
\be\ba{lll}
\1\B\u\v\C\2
& \localgap \stackrel{{\rm G}i}{=}& \localgap
\1\B(\v\C\2+(-1)^c \2\C^\dagger\v)\u-(-1)^c \1\B\u\2\C^\dagger\v
\\

& \localgap =& \localgap\phantom{-}
\underline{\1\B\v\C\2\u}
+\1\B\2((-1)^c \C^\dagger\v\u+\u\v\C)\\

&& \localgap
-\underline{\1\B\2\u\v\C}
+\1\B\u\2(-(-1)^c\C^\dagger\v+\v\C)\\

&& \localgap
-\underline{\1\B\u\2\v\C}
\\

& \localgap \stackrel{{\cal I}}{=}& \localgap
2\{\1\B\2[\u\v\C]\}
+2\{\1\B\u\2[\v\C]\}
-\1\B\v\C\u\2,

\ea
\label{gi:proof}
\ee
The result consists of the leading terms of two ${\rm R}_{12}$-typed elements
and one ${\rm R}_{12}[*]$-typed element. The leaders of the three terms are lower than $f$.

By (\ref{g3:proof}) and (\ref{gi:proof}), a monomial that is non-reduced with respect to a 
G$i$ for some $i\geq 3$ must contain a subsequence of the form
$\u\D\v$,
where (i) the length of $\D$ is $d>0$; (ii) $\u\succ \v$; (iii) if $\l_\D$ is
the leading variable of $\D$, then $\u\succ \l_\D$; (iv) if $\D$ contains more than one variable, then 
$\l_\D\succ \v$. (\ref{g3:proof}) and (\ref{gi:proof})
can be written in the following unified form:
\be
\u\D\v \stackrel{{\rm G}(d+2)}{=} 
2(\u\v[\D])+2(\v[\u\D])-\D\u\v.
\label{g:reduction}
\ee
It is called the {\it fundamental $\cal I$-reduction formula}.

{\it Step 3}. Consider $\cal I$-reductions on the leader of 
a general ${\rm R}_{12}[*]$-typed element
$f=\1\A\2[\B]-(-1)^a\1\2\A^\dagger[\B]$,
where the length of $\B$ is $b>0$. 

Since $[\B]=(-1)^b[\B^\dagger]$,
henceforth we assume that in any ${\rm R}_{12}[*]$-typed element to be normalized, the leading 
variable of any bracket has higher order than the trailing variable of the bracket. Then the leader
of the bracket is always its representative.

In this step, we consider $\cal I$-reduction to the representative $\B$ of $[\B]$.
Let $[\B]=[\C\u\D\v\E]$.
Substituting (\ref{g:reduction}) into it, we get
\be
[\C\u\D\v\E]\stackrel{{\cal I}}{=}2[\C\u\v\E][\D]+2[\C\v\E][\u\D]-[\C\D\u\v\E].
\label{bracket:reduction}
\ee

{\it Step 4}. Consider $\cal I$-reductions on the leader $\1\A\2\B$ of 
$f=\1\A\2[\B]-(-1)^a\1\2\A^\dagger[\B]$ involving both
the tail part of $\A$ and the head part of $\B$, where
the leading variable of $\B$ is assumed to be higher than the trailing variable.

As $\2$ is lower than any element of $\A, \B$, the only possible reduction is by G3. 
Let $\1\A\2\B=\1\C\a\2\b\D$,
where $\C$ may be empty but $\D$ is not. 
Assume $\a\succ \b\succ \t_\D$, where $\t_\D$ is the trailing variable of $\D$.
Let the length of $\D$ be $d$. It is easy to prove that applying G3 to $\a\2\b$ 
in vector-variable binomial $\1\C\a\2[\b\D]$ is equivalent
to the following {\it absorption of bracket}:
\be\ba{lcl}
\1\C\a\2[\b\D] & \stackrel{{\cal I}}{=} & \1\C[\b\D]\a\2
\\
&=& 2^{-1}(\1\C\b\D\a\2)+(-1)^d 2^{-1}(\1\C\D^\dagger \b\a\2).
\ea
\ee
Each term in the result is a leading term of an ${\rm R}_{12}$-typed element lower than $f$.

{\it Step 5}. In Step 3, we have seen that a single bracket after $\cal I$-reduction, 
may be split into two brackets. The split can continue and we gradually get expressions of the form
\be
{\rm R}_{12}[j]: \ \
 \1\E\2[\F_2][\F_3]\cdots [\F_j]-(-1)^{e}\1\2\E^\dagger[\F_2][\F_3]\cdots [\F_j], 
\ee
where the length of $\E$ is $e>0$, and the length of $\1\E\2\F_2\F_3$ $\cdots \F_j$ is $m$. 
${\rm R}_{12}[j]$ is a $\mathbb K$-linear combination of 
elements of type ${\rm R}_{12}[*]$ if all but one bracket are each expanded into two terms.

Consider $\cal I$-reductions of ${\rm R}_{12}[j]$
involving more than two bracket factors, and $\cal I$-reductions involving $\1\E\2$ and 
more than one bracket factor. Since $\2$ is lower than all elements
of $\E$ and the $\F_i$, G3 is the only possible Gr\"obner base element that may apply to $\2$ and its 
neighbors on both sides simultaneously. G3 can involve only $[\F_2]$ among the brackets.

In $[\F_2][\F_3]\cdots [\F_j]$, 
only G$i$ where $i>3$ can involve more than two brackets. However, since G$i$ is of the form $\u\D\v$ where $\D$ 
is ascending, if the product of the leaders of three brackets is non-reduced with respect to some
G$i$, then the middle bracket must be composed of a subsequence of $\D$ of length $\geq 2$, contradicting with the 
assumption that the leading variable in the middle bracket be higher than the trailing variable.

So each $\cal I$-reduction of ${\rm R}_{12}[j]$
by a single G$i$ where $i\geq 3$, can involve at most two bracket factors, or the $\1\E\2$ and one
bracket factor. 

{\it Step 6}. Consider $\cal I$-reductions on $[\F_1][\F_2]$, where the leading variable in 
each bracket is higher than the trailing variable. If the leading variable $\l_{\F_1}$ of $\F_1$ is higher than
the leading variable $\l_{\F_2}$ of $\F_2$, then an $\cal I$-reduction commuting the two brackets 
reduces the order of their product. Below we always assume $\l_{\F_1}\prec \l_{\F_2}$.

For G3, there are two possibilities to involve both $\F_1, \F_2$ in the leader $\F_1\F_2$: 
two variables at the end of $\F_1$ and the third at the beginning of $\F_2$, or one variable at the end of
$\F_1$ and the other two at the beginning of $\F_2$.
The latter is impossible because $\l_{\F_1}\prec \l_{\F_2}$.
For G$i$ where $i>3$, there are also two possibilities: two variables at the end of 
$\F_1$ and the rest at the beginning of $\F_2$, or one variable at the end of $\F_1$ and the rest 
at the beginning of $\F_2$. The latter is also impossible due to $\l_{\F_1}\prec \l_{\F_2}$.

Case G3. 
Let $[\F_1][\F_2]=[\B\u\v][\w\C\d]$, where $\u\succ \v$, and $\u\succ\w\succ\d$.
Let the length of $\C$ be $c\geq 0$, and let the leading variable of $\B\u$ be $\l$. Then 
$\w\succ \l\succ \v$.
Applying G3 to $\u\v\w$
is equivalent to the following {\it absorption of the second bracket}:
\be\ba{cl}
& [\B\u\v][\w\C\d] 
\\
\stackrel{{\cal I}}{=} & [\B[\w\C\d]\u\v]
\\
=& 2^{-1}[\B\w\C\d\u\v]+(-1)^c 2^{-1}[\B\d\C^\dagger\w\u\v].
\ea
\label{case:g31}
\ee
The leader of each bracket monomial in the result has lower order than the leader of $[\F_1][\F_2]$.

Case G$i$. Let $[\F_1][\F_2]=[\B\u\v][\a\D\w\C]$, where $\a\D$ is ascending, and $\a\succ \u\succ \v\succ \w$.
Let the lengths of $\B, \C, \D$ be $b,c,d$ respectively.
Let the leading variable of $\B\u$ be $\l$, and let the trailing variable of $\w\C$ be $\t$.
Then $\a\succ \l\succ \v$ and $\a\succ \t$.
Applying G$(d+4)$ to $\u\v\a\D\w$, we get
\[\ba{cl}
& 4\,[\B\u\v][\a\D\w\C] 
\\

=& \B\u\underline{\v\a\D\w}\C+(-1)^{c+d}\B\u\underline{\v\C^\dagger\w}\D^\dagger\a \bigstrut\\

& \hfill
+(-1)^b \v\u\B^\dagger\a\D\w\C
+(-1)^{b+c+d}\v\u\B^\dagger\C^\dagger\w\D^\dagger\a
\\

\stackrel{{\cal I}}{=} & \bigstrut
(\v\a\D\w-(-1)^d\w\D^\dagger\a\v)\B\u\C + (-1)^d \B\u\w\D^\dagger\a\v\C\\

& 
+(-1)^{d}((-1)^c \v\C^\dagger\w+\w\C\v)\B\u\D^\dagger\a\\

& 
-(-1)^d \B\u\w\C\v\D^\dagger\a
\\

& 
+(-1)^b \v(\a\D\w\C+(-1)^{c+d}\C^\dagger\w\D^\dagger\a)\u\B^\dagger
\\

=& \v\{\a\D\w(\B\u\C-(-1)^{b+c}\C^\dagger\u\B^\dagger)\bigstrut\\

& \phantom{\v}
+(-1)^b\a\D\w(\C+(-1)^c\C^\dagger)\u\B^\dagger\\

& \phantom{\v}
+(-1)^{c+d} \C^\dagger\w(\B\u\D^\dagger\a+(-1)^{b+d}\a\D\u\B^\dagger)\\

& \phantom{\v}
-(-1)^{b+c}\C^\dagger\w(\a\D-(-1)^d\D^\dagger\a)\u\B^\dagger\}\\

& +(-1)^d \w\{
-\D^\dagger\a\v(\B\u\C-(-1)^{b+c}\C^\dagger\u\B^\dagger)\\

& \phantom{+(-1)^d\v}
-(-1)^{b+c}\D^\dagger\a\v\C^\dagger\u\B^\dagger
-(-1)^{b+d}\C\v\a\D\u\B^\dagger\\

& \phantom{+(-1)^d\v}
+\C\v(\B\u\D^\dagger\a+(-1)^{b+d}\a\D\u\B^\dagger)
\}\\

& +(-1)^d \B\u\w(\D^\dagger\a\v\C-\C\v\D^\dagger\a)
\\

\stackrel{{\cal I}}{=} &\bigstrut
4([\v\a\D\w][\B\u\C]+(-1)^d[\w\C\v][\B\u\D^\dagger\a])\\

&
+(-1)^b\v(\a\D\C+(-1)^{c+d}\D^\dagger\a\C^\dagger)\w\u\B^\dagger\\

&
-(-1)^b\w(\C\v\a\D+(-1)^{c+d}\D^\dagger\a\v\C^\dagger)\u\B^\dagger
\\

& +(-1)^d \B\u\w(\D^\dagger\a\v\C-\C\v\D^\dagger\a)
\\

\ea\]
\be\ba{cl}
\stackrel{{\cal I}}{=} &\bigstrut
4([\v\a\D\w][\B\u\C]+(-1)^d[\w\C\v][\B\u\D^\dagger\a])\\

&
+(-1)^b2\{\v(\a\D[\C]-(-1)^{c}[\a\D]\C^\dagger)\w\u\B^\dagger\}\\

& 
-4\,[\B\u\w][\C\v\a\D]
+ \B\u\w\{\C(\a\D-(-1)^d\D^\dagger\a)\\

& \hfill
+(-1)^d \D^\dagger\a(\C+(-1)^c\C^\dagger)
\}\v
\\

\stackrel{{\cal I}}{=} &\bigstrut
4\{[\v\a\D\w][\B\u\C]+(-1)^d[\w\C\v][\B\u\D^\dagger\a]\\

& \phantom{4}
-[\B\u\w][\C\v\a\D]
+ [\B\u\w(\C[\a\D]+(-1)^d \D^\dagger\a[\C])\v]
\}
\\

\stackrel{{\cal I}}{=} &\bigstrut
4\{[\v\a\D\w][\B\u\C]+(-1)^d[\w\C\v][\B\u\D^\dagger\a]\\

& \phantom{4}
-[\B\u\w][\C\v\a\D]
+ [\B\u\w\C\v][\a\D]\\

& \phantom{4}
+(-1)^d[\B\u\w\D^\dagger\a\v][\C]
\}.
\ea
\label{proof:long}
\ee
The leader of each bracket monomial in the result has lower order than the leader of $[\F_1][\F_2]$.

{\it Step 7}. Consider a bracket of the form $h=[\a_1\B_1\c_1\a_2\B_2\c_2$ $
\cdots\a_k\B_k\c_k]$, where
(1) $k>1$, (2) $\a_i\succ \c_i$ for every $i$, (3) 
$\a_1\succ \c_j$ for all $1\leq j\leq k$. Let the length of $\B_i$ be $b_i$.

When $k=2$, 
\[\ba{cl}
& \localgap
[\a_1\B_1\c_1\a_2\B_2\c_2]-2[\a_1\B_1\c_1][\a_2\B_2\c_2]
\\

= &\bigstrut\localgap
2^{-1}\{
(-1)^{b_1+b_2}\c_2\B_2^\dagger\a_2\c_1\B_1^\dagger\a_1
-(-1)^{b_2}\underline{\a_1\B_1\c_1\c_2\B_2^\dagger\a_2}
\\

& \hfill
-(-1)^{b_1}\c_1\B_1^\dagger\a_1\a_2\B_2\c_2
-(-1)^{b_1+b_2}\underline{\c_1\B_1^\dagger\a_1\c_2\B_2^\dagger\a_2}
\}
\\

\stackrel{{\cal I}}{=} &\localgap \bigstrut
-(-1)^{b_2}[\c_2\B_2^\dagger\a_2\a_1\B_1\c_1].
\ea
\]
The leader in the result has lower order than $h$. 

For $k>2$, 
\[\ba{cl}
& [\a_1\B_1\c_1\cdots \a_k\B_k\c_k]\\
& \hskip .6cm -2[\a_1\B_1\c_1\cdots \a_{k-1}\B_{k-1}\c_{k-1}][\a_k\B_k\c_k]
\\
\stackrel{{\cal I}}{=} &
-(-1)^{b_k}[\c_k\B_k^\dagger\a_k(\a_1\B_1\c_1\cdots \a_{k-1}\B_{k-1}\c_{k-1})],
\ea
\]
and for $[\a_1\B_1\c_1\cdots \a_{k-1}\B_{k-1}\c_{k-1}]$, the split into 
$[\a_1\B_1\c_1$ $\cdots \a_{k-2}\B_{k-2}\c_{k-2}][\a_{k-1}\B_{k-1}\c_{k-1}]$
can continue. In the end, we get
\be
[\a_1\B_1\c_1\cdots \a_k\B_k\c_k]
\stackrel{{\cal I}}{=} 2^{k-1}[\a_1\B_1\c_1]\cdots [\a_k\B_k\c_k]
+g,
\label{split:form}
\ee
where $g$ is a bracket polynomial whose leader has lower order than $h$.

(\ref{split:form}) can be used to split a long bracket 
whose representative is $\cal I$-normal. It can also be used in the converse direction,
to concatenate short brackets into a long one.

{\it Step 8}. So far we have proved that for any ${\rm R}_{12}$-typed element 
$\1[\E\2]$ or any ${\rm R}_{12}[j]$-typed element
$\1[\A\2][\F_2][\F_3]\cdots[\F_j]$, as long as the
leader is not $\cal I$-normal, 
$\cal I$-reductions can always be carried out to change $\1\E\2$ or 
$\1\A\2[\F_2][\F_3]\cdots[\F_j]$ into the following form:
\be\ba{l}
\ds \1 T=\sum_\alpha \lambda_\alpha\1\E_\alpha\2
+\sum_{\beta} \mu_{\beta}\1\A_{\beta}\2[\F_{\beta_2}][\F_{\beta_3}]\cdots [\F_{\beta_j}]\\

\ds \hfill
+\sum_\gamma \tau_\gamma \1\2[\D_{\gamma_1}][\D_{\gamma_2}]\cdots [\D_{\gamma_k}],
\ea
\label{t:end}
\ee
where the leading variable in each bracket is higher than the trailing variable, the
$\1\E_\alpha\2$ and
$\1\A_{\beta}\2\F_{\beta_2}\F_{\beta_3}\cdots \F_{\beta_j}$ 
are all $\cal I$-normal.

Since any $\cal I$-normal form is of type $\Y_1\z_1\ldots \Y_k\z_k$ or 
$\Y_1\z_1$ $\ldots\Y_k\z_k\Y_{k+1}$, it must be that \\
(i) $\1\E_\alpha\2=\Y_1\z_1$,
{\it i.e.}, $\E_\alpha=\3\4\cdots \m$. \\
(ii) 
$\1\A_{\beta}\2\F_{\beta_2}\F_{\beta_3}\cdots \F_{\beta_j}
=\Y_1\z_1\ldots \Y_{j}\z_{j}$, and 
\[\left\{\ba{cll}
\1\A_{\beta}\2 &=& \Y_1\z_1,\\
\F_{\beta_2} &=& \Y_2\z_2,\\
\ldots && \ldots \\
\F_{\beta_j} &=& \Y_{j}\z_{j}.
\ea\right.\]

(i) is obvious. In (ii), the trailing variable of each bracket must be some $\z_i$.
If an $\F_{\beta_i}$ is $\Y_{h}\z_{h}\ldots \Y_{h+p}\z_{h+p}$ for some $p>0$, for $h\leq s\leq h+p$,
let the
leading variable of $\Y_s$ be $\l_s$, then $\l_h\succ \z_{h+p}\succ \z_{h+p-1}\succ \ldots \succ \z_{h}$. 
By (\ref{split:form}), $[\F_{\beta_i}]$ is split into $2^{p}[\Y_{h}\z_{h}]\ldots [\Y_{h+p}\z_{h+p}]$
plus some bracket monomials of lower leader. Then $\cal I$-reductions continue to the 
terms involving such bracket monomials. Ultimately each bracket is of the form
$[\Y_i\z_i]$.

In (\ref{t:end}), $\1\E_\alpha\2=(-1)^m \1(\E_\alpha\2)^\dagger$ by ${\rm R}_{12}^N[1]$, 
$\1\A_{\beta}\2$ $[\F_{\beta_2}]\cdots$ $[\F_{\beta_j}]
=(-1)^m \1(\A_{\beta}\2[\F_{\beta_2}]\cdots [\F_{\beta_j}])^\dagger$ by ${\rm R}_{12}^N[j]\strut$,
and $\1\2[\D_{\gamma_1}]$ $\cdots [\D_{\gamma_k}]
=(-1)^m\1(\2[\D_{\gamma_1}]\cdots [\D_{\gamma_k}])^\dagger$
by ${\cal I}[{\cal M}]$.  So $\1 T-(-1)^m \1 T^\dagger$ is reduced to zero by ${\cal I}[{\cal M}]$
and ${\rm R}_{12}^N$. 
\endproof

\section{Square-free vector-variable \\
polynomial ring}
\setcounter{equation}{0}
\vskip .2cm

When $\cal M$ is a general multiset of vector variables,
a {\it square} in ${\cal Q}[{\cal M}]$
refers to the product of a vector with itself.
Denote $\v_i^2:=\v_i\v_i$. It commutes with everything in ${\cal Q}[{\cal M}]$.

\bp \label{squarefree}
In ${\cal Q}[{\cal M}]$,
let $f=g\v_i^2h$ be a multiplier of $\v_i^2$. If $f$ is not $\cal I$-normal,
then by doing $\cal I$-reduction
to $g,h$, together with rearranging the position of $\v_i^2$ in each term, 
$f$ can become $\cal I$-normal.
\ep

\bproof
Suppose $g, h$ are $\cal I$-normal.
There are three cases for $f$ to be non-reduced with respect to the Gr\"obner base
${\cal G}[{\cal M}]$:

(1)
If $f$ contains as a factor the leader of
G$k$ for $k>3$, or EG$j$ for $j>1$ involving both of $\v_i^2$, then 
$\v_i^2$ is preserved by the reduction with the Gr\"obner base element.

(2)
If $g\v_i$ is non-reduced, then switch the element of ${\cal G}[{\cal M}]$ with respect to
which $g\v_i$ is non-reduced:

Case EG2: Let $g\v_i=\A\u\v_i\v_i$ or $\A\u\u\v_i$, where $\u\succ \v_i$. Then
$\A\u\v_i\v_i^2h=\A\v_i^2\u\v_ih$ or $\A\u\u\v_i^2h=\A\v_i^2\u\u h$.
\\
Case G3: Let $g\v_i=\A\u\w\v_i$ where $\u\succ \w$ and $\u\succ \v_i$. Then
$\A\u\w\v_i^2h=\A\v_i^2\u\w h$.
\\
Case G$k$ for $k>3$ or EG$j$ for $j>2$: Let $g\v_i=\C\u\w\D\v_i$ where $\u\succ \w\succ \v_i$. Then
$\C\u\w\D\v_i^2h=\C\v_i^2\u\w\D h$.

(3)
If $\v_ih$ is non-reduced, then switch the element of ${\cal G}[{\cal M}]$ with respect to
which $\v_ih$ is non-reduced:

Case EG2: Let $\v_ih=\v_i\v_i\z\B$ or $\v_i\z\z\B$, where $\v_i\succ \z$. Then
$g\v_i^2\v_i\z\B=g\v_i\z\v_i^2\B$ or $g\v_i^2\z\z\B=g\z\z\v_i^2\B$.
\\
Case G3: Let $\v_ih=\v_i\y\z\B$ where $\v_i\succ \y$ and $\v_i\succ \z$. Then
$g\v_i^2\y\z\B=g\y\z\v_i^2\B$.
\\
Case G$k$ for $k>3$ or EG$j$ for $j>2$: Let $\v_ih=\v_i\y\C\z\D$ where $\v_i\succ \y\succ \z$. Then
$g\v_i^2\y\C\z\D=g\y\v_i^2\C\z\D$.

In all the cases, the order of $f$ is decreased while preserving $\v_i^2$. 
By induction on the order we get the
conclusion.
\endproof

Proposition \ref{squarefree} suggests a 
``square-free normalization" of vector-variable polynomials, by moving all squares to a set 
free of any reduction operation, and maintaining the set of squares in a normal form.

In a vector-variable monomial, let the set of squares be separated from the remainder of the monomial
by a symbol ``$\square$", such that all elements on the right side of the symbol are squares. 
Two things need to be established before such a symbol can be used in algebraic manipulations:
(1) algebraic structure of the new symbolic system, (2) connection with the canonical system
based on V2, V3, V4. 

Let $\cal S$ be a commutative monoid. All elements in $\cal S$ span a $\mathbb K$-vector space
whose dimension equals the number of elements in $\cal S$. The product in the vector space is
the multilinear extension of the product in $\cal S$. The vector space equipped with this product
forms a commutative $\mathbb K$-algebra, called the {\it $\mathbb K$-algebra extension} of monoid $\cal S$, 
denoted by ${\mathbb K}{\cal S}$.

For a $\mathbb K$-algebra $\cal A$, when $\cal S$ is a subset of the center of $\cal A$, then
$\cal A$ is not only a module over the ring ${\mathbb K}{\cal S}$, but a multilinear algebra
over ${\mathbb K}{\cal S}$, called a {\it ${\mathbb K}{\cal S}$-algebra}.

For $\mathbb K$-tensor algebra $\bigotimes[{\cal M}]$,
let 
\be
\hbox{$\bigotimes$}^\square[{\cal M}]:=\hbox{$\bigotimes$}[{\cal M}]/\langle {\rm V2}\rangle. 
\ee
It is easy to see that when setting ${\cal S}$ to be generated by elements of the form
$\v_i^2:=\v_i\otimes \v_i$, for all $\v_i\in {\cal M}$, then ${\bigotimes}^\square[{\cal M}]$ is 
a ${\mathbb K}{\cal S}$-algebra, called the {\it ${\mathbb K}{\cal S}$-tensor algebra}
over multiset $\cal M$, or the {\it square-free tensor algebra} over $\cal M$.
The product in ${\bigotimes}^\square[{\cal M}]$ is induced from the tensor product. For
brevity we still denote the product by ``$\otimes$", but denote the commutative product in $\cal S$ by
juxtaposition of elements.

In ${\bigotimes}^\square[{\cal M}]$, for all $q\in \bigotimes[{\cal M}]$ and $s\in {\cal S}$,
we introduce the notations
\be\ba{llll}
q\square s &:=& q\otimes s & \in \bigotimes[{\cal M}],\\
q\square &:=& q & \in \bigotimes[{\cal M}],\\
\square s &:=& s & \in {\cal S}.
\ea
\ee
Then
\be
(q_1\square s_1)\otimes(q_2\square s_2):=q_1\otimes q_2\square s_1 s_2.
\ee

Formally, an element $q\in {\cal Q}[{\cal M}]$ is taken as $q\square 1$, and
an element $s\in {\mathbb K}{\cal S}$ is taken as $1\square s$. In other words,  
factor $\square 1$ (or $\square$) in $q\square 1$ (or $q\square$) is usually omitted. So
\be
q\square s=q\otimes (\square s)=(\square s)\otimes q,
\ee
and
\be
\square s t=(\square s)\otimes(\square t)=(\square t)\otimes(\square s).
\ee

That $\cal S$ is generated by squares can be succinctly expressed by the following identity:
\be 
\v_i\otimes \v_i
= \square \v_i^2.
\label{def:square}
\ee
making left multiplication with $f$ and right multiplication with $g$ on both sides of the identity, we get
$f\otimes (\v_i\otimes \v_i)\otimes g
=f\otimes g \square \v_i^2$. It includes V2 as a special case.

The {\it degree}, or {\it length}, of a monomial in $\bigotimes^\square[{\cal M}]$ is the
degree of the monomial when taken as an element in $\bigotimes[{\cal M}]$.
The {\it left degree} or {\it left length} of
a monomial refers to the degree of the monomial on the left side of the square symbol.
For a monomial $f\in \bigotimes^\square[{\cal M}]$, its {\it canonical form} in 
$\bigotimes[{\cal M}]$ is defined to be the monomial of lowest 
lexicographic order among all monomials equal to $f$ modulo V2. The {\it order} of $f$ is that of its
canonical form. This ordering is still called the {\it lexicographic ordering}.

The canonical form of $f=\v_{i_1}\otimes \v_{i_2}\otimes \cdots \otimes \v_{i_k}
\square \v_{j_1}^{2r_1}\v_{j_2}^{2r_2}$ $\cdots \v_{j_l}^{2r_l}$,
where $\v_{j_1}\prec \v_{j_2}\prec\ldots\prec \v_{j_l}$,
can be obtained as follows: \\
1. Set $g=\v_{i_1}\otimes \v_{i_2}\otimes \cdots \otimes \v_{i_k}$.\\
2. For $p$ from 1 to $l$, let $\v_{i_t}$ be the first variable in the sequence of $g$ such that
$\v_{i_t}\succ \v_{j_p}\succeq \v_{i_{t-1}}$. Insert 
\[
{\underbrace{\v_{j_p}\otimes \v_{j_p}\otimes\cdots \otimes \v_{j_p}}_{2r_p}}
\] 
to the position before $\v_{i_t}$ in $g$,
and update $g$.\\
3. Output $g$.

The vector-variable polynomial ring ${\cal Q}[{\cal M}]$ when taken as the quotient of
$\bigotimes^\square[{\cal M}]$ modulo the two-sided ideal ${\cal I}^\square[{\cal M}]$ generated by
V3, V4, is a ${\mathbb K}{\cal S}$-algebra, called the 
{\it square-free  polynomial ring}, denoted by 
${\cal Q}^\square[{\cal M}]$. The product in ${\cal Q}^\square[{\cal M}]$ is still denoted by 
juxtaposition of elements.  ${\cal I}^\square[{\cal M}]$ is called the {\it syzygy ideal} of
${\cal Q}^\square[{\cal M}]$. 

Theorem \ref{main:2} has the following square-free version for multiset $\cal M$:

\bt \label{main:2square}
Let ${\cal M}$ be a multiset of $m$ symbols, among which $n$ are different ones: 
$\v_1\prec \v_2\prec\ldots\prec \v_n$, and let 
${\cal I}^\square[{\cal M}]$ be the syzygy ideal of the square-free  polynomial ring 
${\cal Q}^\square[{\cal M}]$. 

(1) [Gr\"obner base] The following are a reduced Gr\"obner base of ${\cal I}^\square[{\cal M}]$:
$\rm G3$, and
\bu
\item[${\rm G}j$:] for all $3< j< n+1$, and $i_1<i_2<\ldots< i_j$,
\[
{[}\v_{i_3}\v_{i_2}\v_{i_4}\v_{i_5}\cdots \v_{i_j}\v_{i_1}]
-{[}\v_{i_2}\v_{i_4}\v_{i_5}\cdots \v_{i_j}\v_{i_1}\v_{i_3}],
\]

\item[${\rm EG}k$:] for all $3\leq k\leq n+1$, and $i_1<i_2<\ldots< i_k$,
\[
{[}\v_{i_3}\v_{i_2}\v_{i_3}\v_{i_4}\cdots \v_{i_k}\v_{i_1}]
-{[}\v_{i_2}\v_{i_3}\v_{i_4}\cdots \v_{i_k}\v_{i_1}\v_{i_3}].
\]
\eu
The above Gr\"obner base is denoted by ${\cal G}^\square[{\cal M}]$.

(2) [Normal form] In a normal form, every term is up to coefficient
of the form $\V_{Y_1}\v_{z_1}\V_{Y_2}\v_{z_2}\cdots \V_{Y_k}\v_{z_k}\square s$ or \\
$\V_{Y_1}\v_{z_1}\cdots \V_{Y_k}\v_{z_k}\V_{Y_{k+1}}\square s$,
where \\
(i) $k\geq 0$, \\
(ii) $\v_{z_1}\v_{z_2}\cdots \v_{z_k}$ is non-descending, \\
(iii) 
every $\V_{Y_i}$ is an ascending monomial of length $> 0$, \\
(iv) 
$\V_{Y_1}\V_{Y_2}\cdots \V_{Y_k}$ 
(or $\V_{Y_1}\cdots \V_{Y_k}\V_{Y_{k+1}}$
if $\V_{Y_{k+1}}$ occurs) is non-descending, 
\\
(v) for every $i\leq k$, let $\v_{t_i}$ be the trailing variable of $\V_{Y_i}$, then
$\v_{t_i}\succ \v_{z_i}$,
\\
(vi) $s$ is either 1 or the product of several squares.
\et

In ${\cal Q}^\square[{\cal M}]$, 
a  polynomial is said to be {\it ${\cal I}^\square$-normal} 
if its leading term is reduced with respect to
the Gr\"obner base ${\cal G}^\square[{\cal M}]$. 

A monic {\it square-free uni-bracket monomial} is of the form $[\A]\square s$, where 
$\A$ is either 1 or a  monomial of length $>1$, $s$ is a product of squares, and the length
of $\A s$ is $m$.
A {\it square-free uni-bracket polynomial} is a $\mathbb K$-linear combination of
square-free uni-bracket monomial. The space of square-free uni-bracket polynomials is denoted by 
$[{\cal Q}]^{\square}[{\cal M}]$. 

The $\mathbb K$-linear space of degree-$m$ square-free 
polynomials is denoted by ${\cal Q}_m^{\square}[{\cal M}]$.
The space $[{\cal Q}]^\square[{\cal M}]$ can be taken as the quotient of ${\cal Q}_m^{\square}[{\cal M}]$
modulo the ideal 
\be
{\cal J}^\square[{\cal M}]:=
{\cal I}^\square[{\cal M}]+[{\cal I}]^\square[{\cal M}],
\ee
where 
$[{\cal I}]^{\square}[{\cal M}]$ is
composed of the vector parts of degree-$m$ square-free
 polynomials, {\it i.e.}, the $\mathbb K$-linear span of elements of the form
\be
{\rm R}^\square:\ \ \A\square s-(-1)^{a}\A^\dagger \square s,
\ee
where $\A$ is a monomial of length $a> 0$ and contains no square, $s$ is a product of squares,
and the length of $\A s$ is $m$.

The modulo-$[{\cal I}]^\square[{\cal M}]$ operation identifies a square-free uni-bracket with its 
representative. It removes the outer bracket symbol on the left side of the ``$\square$" symbol from 
every square-free uni-bracket, disregarding the length of the bracket.  
Ideal $[{\cal I}]^\square[{\cal M}]$
is called the {\it uni-bracket removal ideal} in ${\cal Q}^\square[{\cal M}]$, and ideal 
 is called the {\it syzygy ideal} of 
$[{\cal Q}]^\square[{\cal M}]$
in $\bigotimes^\square[{\cal M}]$.

\section{Gr\"obner base and normal form for uni-bracket polynomials}
\setcounter{equation}{0}
\vskip .2cm

In this section, we extend Theorem \ref{main:bracket} to the case of general multiset $\cal M$
with $m\geq 3$ different vector variables. The 
modulo-$[{\cal I}][{\cal M}]$ operation is always assumed,
{\it i.e.}, $[{\cal Q}]^\square[{\cal M}]$ and ${\cal Q}_m^\square[{\cal M}]$
are identical, and a uni-bracket does not have the outer bracket symbol on the 
left side of the ``$\square$" symbol.

The following are elements of ${\cal J}^\square[{\cal M}]$: 
\be\ba{ll}
{\rm R}^\square(k): & (\K-(-1)^k \K^\dagger)\square s, \\

{\rm S}_{1}(k): & (\i\A\b_1\B-\b_1\B\i\A)\square s,\ \hbox{ for } \i\neq \b_1,\\

{\rm S}_{10}(k): & (\i\C\b_1-\b_1\i\C)\square s, \hbox{ for } \i\neq \b_1, \\

{\rm S}_{11}(k): & (\b_1\A\b_1\B-\b_1\B\b_1\A)\square s,\\

{\rm Sq}_{1}(k): & (\b_1\C\b_1-\C\square \b_1^2)\square s,\\

{\rm Sq}_{1}[*](k): & (\b_1\E\b_1-\E\square \b_1^2)[\F]\square s,\\

{\rm R}_1(k): & \b_1[\D]\square s, \\

{\rm R}_{11}(k): & \b_1[\C\b_1]\square s, \\

{\rm R}_{11}[*](k): & \b_1[\E\b_1][\F]\square s, \\

{\rm R}_{12}(k): & \b_1[\C\b_2]\square s, \\

{\rm R}_{12}[*](k): & \b_1[\E\b_2][\F]\square s,
\ea 
\label{basic:square}
\ee
where \\
(a) $s$ has length $m-k$, and the left length of each expression is $k>0$;\\
(b) $\b_1, \b_2$ are respectively the variables of the lowest 
order and the second lowest order on the left side of the square symbol;
\\
(c) in ${\rm S}_{1}$, either $\A$ or $\B$ can be empty, while in ${\rm S}_{11}$, 
both $\A$ and $\B$ are non-empty; \\
(d) in ${\rm R}_1$, $\D$ is either empty ({\it i.e.}, $\D=1$), or of length $>1$;\\
(e) in ${\rm S}_{10}$ and ${\rm R}_{11}$, $\C$ is non-empty;\\
(f) in ${\rm Sq}_{1}[*](k)$ and ${\rm R}_{11}[*]$, $\E,\F$ are non-empty, and $\F$ does not contain $\b_1$;\\
(g) in ${\rm R}_{12}$, $\C$ is non-empty and does not contain $\b_1$;\\
(h) in ${\rm R}_{12}[*]$, $\E,\F$ are non-empty and do not contain $\b_1$, and
$\F$ does not contain $\b_2$.

In bracket $[\1\A\1]$, we have $[\1\A\1]\stackrel{{\cal I}^\square}{=}[\A]\square \1^2$.
That the leading variable has higher order than the trailing variable is always possible.
This is taken as a postulate for all the brackets in (\ref{basic:square}).

Consider a general element $f=(\K-(-1)^k \K^\dagger)\square s$ of type ${\rm R}^\square(k)$: \\
1.
If $\b_1$ occurs in $\K$ both as the leading variable and trailing variable, then 
$f\in \langle {\rm Sq}_1(k), {\rm R}^\square(k-2)\rangle$. 
\\
2. If $\b_1$ occurs in $\K$ at only one end, then
$f\in \langle {\rm R}_1(k), {\rm S}_1(k)\rangle$. 
\\
3. If $\b_1$ occurs at the interior of $\K$, set $\K=\A\b_1\B$, where $\A, \B$ are both non-empty,
and $\b_1$ does not occur at any end of $\A$ or $\B$.
By (\ref{estab:r}),  $f\in \langle {\rm R}_1(k), {\rm S}_1(k)\rangle$. 

By induction on $k$, we get

\bl
\[
{\rm R}^\square(k)\subseteq 
\sum_{h\leq k} (\langle{\rm S}_1(h)\rangle + \langle{\rm Sq}_1(h)\rangle + \langle{\rm R}_1(h) \rangle).
\]
\el 

In ${\rm R}_1(k)$, when $\D$ contains $\b_1$, let $\D=\A\b_1\B$, where the lengths of $\A, \B$
are respectively $a, k-a-2$, then
\be\ba{lcl}
\b_1[\A\b_1\B]\square s &\stackrel{{\cal I}}{=}& \b_1[\B\A\b_1]\square s \\

&=& (\b_1\B\A\b_1-(-1)^k \A^\dagger\B^\dagger)\square \b_1^2 s \\

&=& (\b_1\B\A\b_1-\B\A\square \b_1^2)\square s\\
&& \hfill
+(\B\A-(-1)^k \A^\dagger\B^\dagger)\square \b_1^2 s \\

&\in & \langle {\rm Sq}_1(k), {\rm R}^\square(k-2)\rangle.
\ea
\label{r11:replace}
\ee
By induction on $k$, we get that both $\b_1[\A\b_1\B]\square s$ and \\
$\b_1[\B\A\b_1]\square s
\in {\rm R}_{11}(k)$
are equivalent to ${\rm Sq}_1(k)$:\
$(\b_1\B\A\b_1-\B\A\square \b_1^2)\square s$ in the sense that their difference is in 
the ideal 
$\sum_{h\leq k-2} (\langle {\rm S}_1(h)\rangle 
+\langle{\rm R}_{11}(h)\rangle
+\langle{\rm R}_{12}(h)\rangle)
+\langle {\rm R}_1(1)\rangle+{\cal I}^\square[{\cal M}]$.

\bl 
\label{observation}
${\rm R}^\square(k)$ is a subset of the ideal 
\[
\sum_{h\leq k} (\langle {\rm S}_1(h)\rangle 
+\langle {\rm R}_{11}(h)\rangle
+\langle{\rm R}_{12}(h) \rangle)
+\langle {\rm R}_1(1)\rangle
+{\cal I}^\square[{\cal M}]. 
\]
\el

\bt 
\label{main:bracket:general}
Let ${\cal M}$ be a multiset of $m>2$ symbols, among which $n\geq 2$ are different ones:
$\1\prec\2\prec \ldots\prec \n$, and let 
${\cal Q}^\square[{\cal M}]$ be the square-free vector-variable polynomial ring over $\cal M$.
Let 
$[{\cal Q}]^\square[{\cal M}]$ be the space of square-free uni-bracket polynomials, and let 
${\cal J}^\square[{\cal M}]$
be its syzygy ideal in $\bigotimes^\square[{\cal M}]$.

(1) [Gr\"obner base] The following are a reduced Gr\"obner base of ${\cal J}^\square[{\cal M}]$, 
denoted by ${\cal B}{\cal G}{[}{\cal M}{]}$: 

\bu
\item[${\cal G}^\square{[}{\cal M}{]}$:] $\rm G$i for all $3\leq i\leq n$;
${\rm EG}j$ for all $3\leq j\leq n+1$.

\item[${\rm S}_1^\square$:] $(\A\b_1\B-\b_1\B\A)\square s$, where the length of each term is $m$,
$\b_1$ is the variable of the lowest order on the left side of the square symbol,
$\A$ is a non-empty ascending sequence not containing $\b_1$, 
and $\A\b_1\B$ is ${\cal I}^\square$-normal;

\item[${\rm R}_{1}^\square{[}j,l{]}$:] 
\be\ba{r}
\b_1[\Y_1\b_1][\Y_2\b_1]\cdots [\Y_j\b_1]
[\Y_{j+1}\z_{j+1}]
[\Y_{j+2}\z_{j+2}]\\
\cdots [\Y_{j+l}\z_{j+l}]\square s,
\ea
\ee
\eu
where (i) $j,l\geq 0$, and the length of each term is $m$,
\\
(ii) $\b_1$ is the variable of the lowest order on the left side of the square symbol,
\\
(iii) each $\Y_i$ is a non-empty ascending sequence not containing $\b_1$,
\\
(iv) for all $j+1\leq i\leq j+l$, $\z_i\neq \b_1$, and 
$\z_i\Y_i$ is ascending;
\\
(v) $\b_1\Y_1\b_1\cdots \Y_j\b_1\Y_{j+1}\z_{j+1}
\cdots \Y_{j+l}\z_{j+l}$ is ${\cal I}^\square$-normal.

\vskip .15cm
(2) [Normal form] In a normal form, every term is ${\cal I}^\square$-normal, and is up to coefficient
of one of the following forms, where statements (ii), (iii) on
$\b_1$ and the $\Y_i$ are still valid: 

\noindent
(I) $\b_1\Y_1\b_1\cdots \Y_j\b_1\Y_{j+1}\z_{j+1}
\cdots \Y_{j+l}\z_{j+l}\Y_{j+l+1}\square s$, where $j,l\geq 0$, and for all $j+1\leq i\leq j+l$,
$\z_i\prec \t_i$, the latter being the trailing variable of $\Y_i$;
\\
(II) $\b_1\Y_1\b_1\cdots \Y_j\b_1\Y_{j+1}\z_{j+1}
\cdots \Y_{j+l}\z_{j+l}\square s$,
where $j,k\geq 0$ but $j+k>0$, each $\z_i\prec \t_i$, but
for some $1\leq h\leq l$, 
if $\l_{j+h}$ is the leading variable of $\Y_{j+h}$, then
$\l_{j+h}\preceq \z_{j+h}$.

\et

{\it Remark}. The set of
${\rm R}_{1}^\square$ can be replaced by the following three sets of degree-$m$ polynomials: for all
$j>0$, $l\geq 0$, 
\be\hskip -.12cm
\ba{ll}
{\rm R}_{1}^\square{[}0,0{]}: &
\b_1\square s, \\

{\rm Sq}_{1}^\square[j,l]: & 
(\b_1\Y_1\b_1-\Y_1\square \b_1^2)[\Y_2\b_1]\cdots [\Y_j\b_1][\Y_{j+1}\z_{j+1}]\bigstrut\\

& \hfill
\cdots [\Y_{j+l}\z_{j+l}]\square s,
\\

{\rm R}_{12}^\square[j,l]: 
& \b_1[\Y_1\b_2][\Y_2\b_2]\cdots [\Y_j\b_2]
[\Y_{j+1}\z_{j+1}][\Y_{j+2}\z_{j+2}]\bigstrut\\
& \hfill 
\cdots [\Y_{j+l}\z_{j+l}]\square s.
\ea
\ee
In ${\rm R}_{12}^\square[j,l]$, $\Y_{j+1}, \ldots, \Y_{j+l}$ do not contain $\b_2$,
and the $\z_i\succ \b_2$. The replacement has no effect upon the normal forms.

\vskip .2cm
\bproof There are several steps.

{\it Step 1}. We need to prove by induction on $k$ that
${\rm S}_1(k)$, \\
${\rm R}_{11}(k), {\rm R}_{12}(k)$
are all in the ideal 
\be
\sum_{{\rm left\, length}\leq k}\langle {\rm S}_1^\square, {\rm R}_1^\square[*]\rangle
+{\cal I}^\square[{\cal M}], 
\label{proof:ideal}
\ee
where the asterisk stands for the $(j,l)$. 

Once this is done, then 
since the leader of 
any element of type ${\rm S}_1^\square$ or
${\rm R}_{1}^\square[*]$ is ${\cal I}^\square$-normal and 
cannot be cancelled by the leader of 
any other element of type ${\rm S}_1^\square$ or ${\rm R}_{1}^\square[*]$, 
the ${\rm S}_1^\square, {\rm R}_{1}^\square[*]$ and ${\cal G}^\square{[}{\cal M}{]}$ must be a 
reduced Gr\"obner base of
$\langle {\rm S}_1,
{\rm R}_{11}, 
{\rm R}_{12}, {\rm R}_{1}(1)\rangle
+{\cal I}^\square[{\cal M}]$. By Lemma \ref{observation}, this ideal is just  
${\cal J}^\square[{\cal M}]$. This proves conclusion (1), and
conclusion (2) follows.

{\it Step 2}. ${\rm S}_1(k)$ requires $k>1$; ${\rm R}_{11}(k)$
and ${\rm R}_{12}(k)$ both require $k>2$. When $k=2$, ${\rm S}_1(2)=(\b_2\b_1-\b_1\b_2)\square s$
is in ${\rm S}_1^\square$. When $k=3$, ${\rm R}_{11}(3)=(\b_1\b_2\b_1-\b_2\square \b_1^2)\square s
={\rm R}^\square_1[1,0]$, and ${\rm R}_{12}(3)=\b_1[\b_3\b_2]\square s={\rm R}^\square_1[0,1]$, where
$\b_3\succ \b_2$.

Consider ${\rm S}_1(3)$. 
There are 3 elements led by variable $\b_2$: $(\b_2\b_1\b_2-\b_1\square \b_2^2)\square s$,
$(\b_2\b_1\b_3-\b_1\b_3\b_2)\square s$, and $(\b_2\b_3\b_1-\b_1\b_2\b_3)\square s$. 
They all belong to ${\rm S}_1^\square$.
There are two other elements in ${\rm S}_1(3)$: $(\b_3\b_1\b_2-\b_1\b_2\b_3)\square s$ and 
$(\b_3\b_2\b_1-\b_1\b_3\b_2)\square s$. By
\[\ba{lcl}
\b_3\b_2\b_1-\b_1\b_3\b_2 &\stackrel{{\rm G}3}{=}& \b_1\b_2\b_3-\b_2\b_3\b_1;
\\
\b_3\b_1\b_2-\b_1\b_2\b_3 &\stackrel{{\rm G}3}{=}& 
(\b_2\b_3\b_1-\b_1\b_2\b_3)\\
&& +(\b_2\b_1\b_3-\b_1\b_3\b_2),
\ea
\]
both are in $\langle {\rm S}_1^\square \rangle+{\cal I}^\square[{\cal M}]$.

So the statement in Step 1 holds for $k\leq 3$. Assume that it holds for all $k<h$.
When $k=h$, we need to make ${\cal I}^\square$-reduction to the leaders of the
elements of any of the types
\be
{\rm S}_1(h), \
{\rm R}_{11}(h), \
{\rm R}_{12}(h), \
{\rm R}_{11}[*](h), \
{\rm R}_{12}[*](h), 
\label{types}
\ee
at the same time keep the reduction result to be within the 
$\mathbb K$-linear space spanned by elements of the types listed in
(\ref{types}) but where the left length $h$ is replaced by all $i\leq h$. 
Then ultimately all the leaders of these elements
become ${\cal I}^\square$-normal.

{\it Step 3}. Consider types ${\rm R}_{11}(h), {\rm R}_{11}[*](h), {\rm R}_{12}(h),
{\rm R}_{12}[*](h)$. 
Let there be an ${\rm R}_{11}(h)$-typed element $f=2(\b_1[\A\b_1]\square s)$,
and an ${\rm R}_{12}(h)$-typed element $g=2(\b_1[\B\b_2]\square s)$, 
where $\B$ does not contain $\b_1$.
In the following we omit the factor $\square s$. 

Do ${\cal I}^\square$-reductions to $\A,\B$, and assume that
the results are
\be\ba{lcl}
\A&\stackrel{{\cal I}^\square}{=}&\C_\A\b_1+\b_1\D_\A+\b_1\E_\A\b_1+\A^N, \\
\B&\stackrel{{\cal I}^\square}{=}&\C_\B\b_2+\b_2\D_\B+\b_2\E_\B\b_2+\B^N,
\ea
\label{reduction:ab}
\ee
where  
(i)
none of the terms in 
$\C_\A, \D_\A, \E_\A, \A^N$ has $\b_1$ at any end; \\
(ii)
none of the terms in 
$\C_\B, \D_\B, \E_\B, \B^N$ has $\b_2$ at any end;\\
(iii) any of the four terms in each result may not occur;\\
(iv) the component on the right side of the square symbol in each term, together with the symbol itself, are omitted,
as they do not affect the analysis below;\\
(v) in the extreme case, $\A^N$ or $\B^N$ may be in $\mathbb K$, if all vector variables 
in the term form squares and are moved to the right side of the square symbol.

Substituting the reduction results into $f,g$, we get
\[\ba{lcll}
f
&\stackrel{{\cal I}^\square}{=}&\phantom{-}
(\D_\A\b_1-(-1)^h\D_\A^\dagger\b_1) \square \b_1^2
& \hskip .2cm \in \langle {\rm S}_{1}(h-2),\ \\
&&& \hfill {\rm R}_{1}(h-2)\rangle
\\

&& +(\b_1\C_\A-(-1)^h \b_1\C_\A^\dagger)\square \b_1^2
& \hskip .2cm \in {\rm R}_{1}(h-2) 
\\

&& +(\E_\A\square \b_1^2-(-1)^h \b_1\E_\A^\dagger \b_1)\square \b_1^2
& \hskip .2cm \in {\rm R}_{11}(h-2) 
\\

&& +2(\b_1[\A^N\b_1]),
& \hskip .2cm \in {\rm R}_{11}(h)
\\

g
&\stackrel{{\cal I}^\square}{=}&\phantom{-}
2(\b_1[\C_\B]) \square \b_2^2
& \hskip .2cm \in {\rm R}_{1}(h-2)
\\

&& 
+2(\b_1[\D_\B]) \square \b_2^2
& \hskip .2cm \in {\rm R}_{1}(h-2)
\\

&& +2(\b_1[\E_\B \b_2])\square \b_2^2
& \hskip .2cm \in {\rm R}_{12}(h-2)
\\

&& +2(\b_1[\B^N\b_2]).
& \hskip .2cm \in {\rm R}_{12}(h)
\ea
\]
Notice that the left lengths indicated on the right column are the maximal possible ones
for the corresponding types. So
by induction hypothesis, we can assume that in $f,g$, monomials $\A=\A^N, \B=\B^N$ 
and both are $\is$-normal.

The $\is$-reduction to the leaders of $f,g$ are much the same with the procedure
in the proof of Theorem \ref{main:bracket} starting from Step 3 there to Step 8,
with negligible revisions. Formula (\ref{split:form}) can also be used to split the leader
of factor $\b_1[\Y_1\b_1$ $\cdots \Y_k\b_1]$ in a type-${\rm R}_{11}[*]$ element, 
and the leader of factor
$\b_1[\Y_1\b_2\cdots \Y_k\b_2]$ in a type-${\rm R}_{12}[*]$ element.

By induction on the order of the leader, we get that
${\rm R}_{11}(h)$, ${\rm R}_{12}(h), {\rm Sq}_{1}(h), {\rm Sq}_{1}[*](h),
{\rm R}_{11}[*](h)$, 
${\rm R}_{12}[*](h)$
are all in (\ref{proof:ideal}) where $k=h$.

{\it Step 4}. Consider a general type-${\rm S}_{10}(h)$ element $f=(\A\b_1-\b_1\A)\square s$.
Let the ${\cal I}^\square$-reduction result of $\A$ be as in (\ref{reduction:ab}). Then 
if omitting ``$\square s$", 
\[\ba{lcll}
f
&\stackrel{{\cal I}^\square}{=}& \phantom{-}
\C_\A\square \b_1^2-\b_1\C_\A\b_1
& \hskip .2cm \in  {\rm Sq}_{1}(h)\\

&& +\b_1\D_\A\b_1-\D_\A\square \b_1^2
& \hskip .2cm \in {\rm Sq}_{1}(h) 
\\

&& +(\b_1\E_\A-\E_\A \b_1)\square \b_1^2
& \hskip .2cm \in {\rm S}_{1}(h-2) 
\\

&& +\A^N\b_1-\b_1\A^N.
& \hskip .2cm \in {\rm S}_{10}(h)
\ea
\]
So we can assume that in $f=(\A\b_1-\b_1\A)\square s$, monomial $\A=\A^N$ and is $\is$-normal.

The $\is$-reduction to the leading term of $f$ is much the same with the procedure
in the proof of Lemma \ref{lem:s} starting from Case (i) there to Case (ii). 
By induction on the order of the leading term, we get that ${\rm S}_{10}(h)$ is 
in (\ref{proof:ideal}) where $k=h$.

{\it Step 5}. Consider a general type-${\rm S}_{1}(h)$ element $g=(\A\b_1\B-\b_1\B\A)\square s$, 
where $\A, \B$ are both non-empty.
Let the ${\cal I}^\square$-reduction results of $\A, \B$ be as in (\ref{reduction:ab}),
where every $\b_2$ is replaced by $\b_1$. Then if omitting ``$\square s$", 
\[\ba{lcl}
g
&\stackrel{{\cal I}^\square}{=}& \phantom{-}
\C_\A\C_\B\b_1\square \b_1^2-\b_1\C_\B\b_1\C_\A\b_1
\\
&&
+(\C_\A\b_1\D_\B-\D_\B\C_\A\b_1)\square \b_1^2
\\
&& 
+(\C_\A\b_1\E_\B\b_1-\E_\B \b_1\C_\A\b_1)\square \b_1^2
\\
&&
+\C_\A\B^N\square \b_1^2-\b_1\B^N\C_\A\b_1
\\

&&
+\b_1\D_\A\b_1\C_\B\b_1-\b_1\C_\B\D_\A\square \b_1^2
\\
&&
+(\b_1\D_\A\D_\B-\D_\B\b_1\D_\A)\square \b_1^2
\\
&&
+\b_1\D_\A\E_\B\b_1\square \b_1^2-\E_\B\D_\A\square \b_1^4
\\
&&
+\b_1\D_\A\b_1\B^N-\b_1\B^N\b_1\D_\A
\\

&&
+(\b_1\E_\A\C_\B\b_1-\b_1\C_\B\E_\A\b_1)\square \b_1^2
\\
&&
+(\b_1\E_\A\b_1\D_\B-\D_\B\b_1\E_\A\b_1)\square \b_1^2
\\
&&
+\b_1\E_\A\b_1\E_\B\b_1\square \b_1^2-\E_\B\E_\A\b_1\square \b_1^4
\\
&&
+\b_1\E_\A\B^N\b_1^2-\b_1\B^N\b_1\E_\A\b_1
\\

&&
+\A^N\b_1\C_\B\b_1-\b_1\C_\B\b_1\A^N
\\
&&
+(\A^N\D_\B-\D_\B\A^N)\square \b_1^2
\\
&&
+\A^N\b_1\E_\B\b_1-\b_1\E_\B\b_1\A^N
\\
&&
+\A^N\b_1\B^N-\b_1\B^N\A^N.
\ea
\]
In the above result, the lines that do not belong to the ideal
$\langle {\rm Sq}_1(h), {\rm S}_1(h-2), {\rm Sq}_1(h-2), {\rm S}_1(h-4) \rangle$
are
\be\ba{ll}
\phantom{-}
\b_1\D_\A\b_1\B^N-\b_1\B^N\b_1\D_\A & \hskip .2cm \in \langle {\rm S}_{11}(h), {\rm Sq}_1(h)\rangle
\\
+\A^N\b_1\C_\B\b_1-\b_1\C_\B\b_1\A^N & \hskip .2cm\in \langle {\rm S}_{11}(h),  {\rm S}_{10}(h)\rangle
\\
+\A^N\b_1\E_\B\b_1-\b_1\E_\B\b_1\A^N & \hskip .2cm\in \langle {\rm S}_{11}(h),  {\rm S}_{10}(h)\rangle
\\
+\A^N\b_1\B^N-\b_1\B^N\A^N. & \hskip .2cm\in {\rm S}_1(h)
\ea
\label{result:lines}
\ee

Some remarks on (\ref{result:lines}) are necessary. 
The first line of (\ref{result:lines}), if nonzero, is ${\rm S}_{11}(h)$ when 
$\B^N\notin \mathbb K$, and ${\rm Sq}_1(h)$ otherwise. By 
$\A^N\b_1\C_\B\b_1-\b_1\C_\B\b_1\A^N=(\A^N\b_1\C_\B\b_1-\b_1\A^N\b_1\C_\B)
+(\b_1\A^N\b_1\C_\B-\b_1\C_\B\b_1\A^N)$, the second line of (\ref{result:lines}) 
is a $\mathbb K$-linear combination of an element of type ${\rm S}_{10}(h)$ and another element of
type ${\rm S}_{11}(h)$.

Consider a general type-${\rm S}_{11}(h)$ element
$p=(\b_1\A\b_1\B-\b_1\B\b_1\A)\square s$, where the length of $\A$ is $a>0$.
When omitting $\square s$, 
\[\ba{lcl}
p &\localgap\stackrel{\is}{=} & \localgap
\b_1\B(\b_1\A-(-1)^a\A^\dagger\b_1)\\

&& 
+(-1)^a\A^\dagger\B\square \b_1^2-\b_1\B\b_1\A\\

&\localgap=& \localgap\bigstrut
(-1)^a(\A^\dagger\B\square \b_1^2-\b_1\B\A^\dagger\b_1)\\

&\localgap\stackrel{{\rm Sq}_1(h)}{=}& \localgap
(-1)^a(\A^\dagger\B-\B\A^\dagger)\square \b_1^2\\

&\localgap\in& \localgap
\langle {\rm S}_1(h-2)\rangle.\bigstrut
\ea
\]
So ${\rm S}_{11}(h)$ is in
(\ref{proof:ideal}) where $k=h$.

By (\ref{result:lines}), we can assume that in $g=(\A\b_1\B-\b_1\B\A)\square s$, 
monomials $\A=\A^N, \B=\B^N$ and both are $\is$-normal.
The $\is$-reduction to the leading term of $g$ is much the same with that
in the proof of Lemma \ref{lem:s} starting from Case (i) there to Case (iii).
By induction on the order of the leading term, we get that ${\rm S}_{1}(h)$ is 
in (\ref{proof:ideal}) where $k=h$.
\endproof

Consider a bracket polynomial $f$ whose multiset of variables is $\cal M$. 
In any term of $f$, when all the brackets but one are expanded into two terms by definition, 
$f$ is changed into a uni-bracket polynomial $g$. Using the Gr\"obner base ${\cal BG}[{\cal M}]$
to make reduction to $g$ results in a uni-bracket polynomial $h$, where
each term is ${\cal I}^\square$-normal. $h$ must have the lowest
order lexicographically among
all uni-bracket polynomials equal to $f$. It
is called the {\it lowest-representative normal form} of $g$, or the 
{\it uni-bracket normal form} of $f$.

{\it Remark}. In the above definition of normal forms, we only considered square-free ones.
Of course any normal form can be converted to a canonical one, where all representatives are 
${\cal I}$-normal instead of $\is$-normal. Later on, we consider only square-free ones.

Let the size of $\cal M$ be $m$. Given any partition $(i_1, \ldots, i_k)$ of integer $m$, 
where each $i_j>1$,
there is a {\it Caianiello expansion} \cite{li} of uni-bracket polynomials into bracket polynomials 
where each term is composed of $k$ brackets of length $i_1, \ldots, i_k$ respectively.
Each expansion produces a normal form. 
It is not clear if such a normal form is of any value.

\section{Normalization of bracket \\
polynomials}
\setcounter{equation}{0}
\vskip .2cm

In this section, we do NOT remove bracket symbols from uni-brackets in 
multiset of variables $\cal M$.

Consider attaching an additional vector variable $\v_0\notin \cal M$ to $\cal M$ 
to form a bigger multiset $\tilde{\cal M}$. Let $\v_0\prec $ all variables in $\cal M$.
In the procedure of obtaining ${\rm R}_1^\square$ by the $\is$-reduction of
$\b_1[\A]$ in the proof of Theorem \ref{main:bracket:general}, or in more details,
in the proof of Theorem \ref{main:bracket}, if the input is 
$\v_0[\A]$ where $\A\in {\cal Q}_{m}[{\cal M}]$, then among the Gr\"obner base of 
${\cal I}^\square[\tilde{\cal M}]$, only those elements in
${\cal G}^\square[{\cal M}]$ are needed in $\is$-reduction.

The reduction of $\v_0[\A]$ is a procedure of recursively 
doing $\is$-reductions to the leader of the bracket polynomial obtained from 
previous $\is$-reductions to $[\A]$. 
At any instance, the representative of a 
bracket monomial in reduction is the leader of the bracket monomial. 
The reduction results in a $\mathbb K$-linear combination of monomials of the form 
$\v_0[\Y_1\z_1]\cdots [\Y_k\z_k]$, where $\z_i\Y_i$ is ascending for every $i$.
After removing $\v_0$ from the result, we get another normal form of uni-bracket 
$[\A]$ in $\cal M$. 

\bt
\label{straightening}
Let $f$ be a square-free bracket polynomial in multiset of variables $\cal M$. 
Do the following to $f$:\\
1. Always select the leader of a bracket as its representative.
\\
2. Rearrange the order of the brackets in the same term, so that 
the leading variables of the brackets are non-descending.
\\
3. Use
(\ref{bracket:reduction}) to normalize the interior of a bracket; it also splits
a bracket into two.
\\
4. Use
(\ref{case:g31}) to absorb a bracket into the one ahead of it.
\\
5. Use (\ref{proof:long}) to decrease the order of the product of two brackets by lifting 
a lower-order variable from the second bracket to the first.
\\
6. Use (\ref{split:form}) to segment a long bracket of type $\Y_1\z_1\Y_2\z_2\cdots$ $\Y_k\z_k$
into short ones.
\\
7.
Once the representative of the leading term of $f$ is $\is$-normal, output the leading term, 
and continue the above
$\is$-normalization to the remainder of $f$.

The output, called the {\it lowest-leader normal form}, or {\it leader-normal form}
is a bracket polynomial where the representative of
each term is up to coefficient of the form
$[\Y_1\z_1][\Y_2\z_2]\cdots$ $[\Y_k\z_k]\square s$, where each $\z_i\Y_i$ is ascending,
$\Y_1\Y_2\cdots \Y_k$ and $\z_1\z_2$ $\cdots \z_k$ are both non-descending, 
and the leading variable $\l_i$
of each $\Y_i$ satisfies $\l_i\succ \z_i$. 
Two bracket polynomials are equal if and only if their leader-normal forms
are identical.
\et

\bproof
After operations 1 and 2, for any bracket monomial in the reduction procedure, its representative
is also its leader, so that the representative of the leading term of bracket polynomial $f$ 
is the leader of $f$.
Once the leader of $f$ is $\is$-normal, it is the leading term of the normal form of vector-variable
polynomial $f$ with respect to the Gr\"obner base ${\cal G}^\square[{\cal M}]$. 
By induction on the order of the output terms from the highest down, we get the uniqueness of the
leader-normal form for $f$.

For two equal bracket polynomials, they have identical uni-bracket normal forms, and so have
identical leader-normal forms.
\endproof

From the above proof, we see that
the leader-normal form of a bracket polynomial $f$ has the following properties: (1)
the representative of any term is the leader of the term;
(2) the representative of the leading term is the leading term of the 
$\is$-normal form of vector-variable polynomial $f$.

By Theorem \ref{main:bracket:general},
the Gr\"obner base ${\cal BG}[\tilde{\cal M}]$ of ${\cal J}^\square[\tilde{\cal M}]$
is composed of ${\cal G}^\square[{\cal M}]$ and the
${\rm R}_1^\square[*]:\, \v_0 g\square s$ for all monomials $g$ in the leader-normal forms of bracket 
polynomials in $\cal M$, such that $gs$ has length $m$. This phenomenon
is easy to understand: the $\mathbb K$-linear subspace of
$[{\cal I}]^\square[\tilde{\cal M}]$ composed of polynomials whose terms are led by variable $\v_0$,
is the space of degree-$(m+1)$ polynomials of the form $\v_0 f\square s$, for all bracket polynomials
$f$ in $\cal M$ such that $fs$ has length $m$. The leader-normal forms are a basis of the 
$\mathbb K$-linear space of length-$m$ square-free bracket polynomials in $\cal M$.

In a leader-normal form, if we commute $\Y_i$ and $\z_i$ in $[\Y_i\z_i]$, we get 
another normal form whose terms are up to coefficient of the form $[\z_1\Y_1]\cdots [\z_k\Y_k]\square s$.
If we write the left side of the square symbol in the following tableau form, where $\Y_i=\y_{i1}\y_{i2}\cdots \y_{it_i}$,
we get
\[
\left[\ba{cccccc}
\z_1 & \y_{11} & \ldots &\ldots & \y_{1t_1} \\
\z_2 & \y_{21} & \ldots & \y_{2t_2}\\
\vdots & \vdots &\ddots \\
\z_k & \y_{k1} & \ldots & \ldots & \ldots & \y_{kt_k}
\ea
\right],
\]
where (1) each row does not need to have equal length, and it is not required that the length be
non-increasing as in Young tableau;\\
(2) each row is an ascending sequence of variables; \\
(3) each column is a non-descending sequence of variables; \\
(4) $\y_{it_i}\preceq \y_{(i+1)1}$ for $1\leq i<k$.

Such a normal form is called the {\it straight form}. Feature (4) above makes this definition stronger than 
the {\it straight form} (or {\it standard form}) of Young tableau. In comparison, in classical bracket 
algebra a bracket monomial is in 
straight form if and only if the entries are ascending along each row, and non-descending along each
column.

The procedure of deriving the straight form of a bracket polynomial is called 
{\it straightening}. Among the formulas used in
Theorem \ref{straightening} for straightening,
(\ref{proof:long}) is highly nontrivial and requires further investigation.

Set $\B\u$, $\a\D$ in (\ref{proof:long}) to be new $\A, \B$ respectively, and let the lengths of 
$\B,\C$ be $b,c$. 
Then (\ref{proof:long}) can be written succinctly as follows:
\be\ba{lll}
{[}\A\v][\B\w\C]
&=& \phantom{-}
[\A\w\C\v][\B]-(-1)^b[\A\w\B^\dagger\v][\C]\\

&& 
-(-1)^b[\w\C\v][\A\B^\dagger]
+(-1)^b[\w\B^\dagger\v][\A\C]\\

&&
-[\A\w][\C\v\B].
\ea
\label{shuffle}
\ee
It is called the {\it shuffle formula for bracket normalization}.

\bp For any two monomials $\A, \B$ of length $a, b$
respectively,
\be\left\{\ba{lll}
\ds\localgap \frac{\A\B+\B\A}{2} &\localgap=& \localgap [\A\B]+(-1)^b(\A[\B^\dagger]-[\A]\B^\dagger), \\

\ds\localgap \frac{\A\B-(-1)^{a+b}\A^\dagger\B^\dagger}{2} &\localgap=&\localgap (-1)^a([\A^\dagger]\B-\A^\dagger[\B]).
\ea\right.
\label{new:reduction}
\ee
\ep

\bproof
\[\ba{ll}
& \A\B+\B\A \\

=& 
2[\A\B]-(-1)^{a+b}\B^\dagger\A^\dagger 
+2[\B]\A-(-1)^b\B^\dagger\A \\

=& 2([\A\B]+[\B]\A-(-1)^b\B^\dagger[\A]).
\ea
\]
\endproof

The fundamental $\cal I$-reduction formula
(\ref{g:reduction}) is a direct consequence of the first identity in (\ref{new:reduction}) for
$\A\B=\u\D$. The shuffle formula (\ref{shuffle}) is a consequence of the following identity by 
making left multiplication with $\A$ and then applying the bracket operator to both sides of the
identity:
\be\ba{lll}
\v[\B\w\C]+\w[\C\v\B]
&=& (-1)^b
(\w\C\v[\B^\dagger]-[\w\C\v]\B^\dagger\\

&&\hfill
-\w\B^\dagger\v[\C]
+[\w\B^\dagger\v]\C).
\ea
\label{shuffle:basic}
\ee
The identity can be obtained as follows: by the second identity 
of (\ref{new:reduction}) from right to left, the right side of (\ref{shuffle:basic}) equals
\[
2^{-1}(
\underline{\B\w\C}\v+\v\underline{\B\w\C})
-(-1)^{b+c}2^{-1}(\underline{\B^\dagger\v\C^\dagger}\w
+\w\underline{\B^\dagger\v\C^\dagger}),
\]
which by the first equality of (\ref{new:reduction}), equals
\[\ba{ll}
& [\v\B\w\C]-(-1)^{b+c}\v[\C^\dagger\w\B^\dagger]\\

&\phantom{[\v\B\w\C]}
-(-1)^{b+c}[\w\B^\dagger\v\C^\dagger]+\w[\C\v\B]\\

=& \v[\B\w\C]+\w[\C\v\B].
\ea
\]

The shuffle formula can be further generalized.
In monomial $[\A\v\D][\B\w\C]$, let the leading variable and trailing variable of any sequence 
$\F$ be $\l_\F$ and $\t_\F$ respectively. Assume
$\l_\A\preceq \l_\B$, and $\l_\A\succ \t_\D$, and $\l_\B\succ \t_\C$.
Further assume $\l_\A\succ \v\succ \w$.
Then $\is$-reductions can be made to $[\A\v\D][\B\w\C]$ to decrease its leader, leading to the following
result:

\bp
Let the lengths of monomials $\A, \B, \C, \D$ be 
$a,b,c,d$ respectively. Then
\be\ba{lll} 
{[}\A\v\D][\B\w\C] &=& \hskip -.1cm
[\v\D\B\w][\A\C]-(-1)^{b+c}[\v\D\C^\dagger\w][\A\B^\dagger]\\

&&
-(-1)^d [\A\w][\D^\dagger\v\B\C]-[\v\D][\A\w\B\C]\\

&& 
-(-1)^b [\A\w\B^\dagger\v\D][\C]
+[\A\w\C\v\D][\B].
\ea
\label{prop:generalf}
\ee
\ep

\bproof
\be\ba{cl}
& 4\,{[}\A\v\D][\B\w\C]
\\

=& \A\underline{\v\D\B\w}\C-(-1)^{b+c}\A\underline{\v\D\C^\dagger\w}\B^\dagger \bigstrut\\

& \hfill
-(-1)^{a+d} \D^\dagger\v\A^\dagger\B\w\C
+(-1)^{a+b+c+d}\D^\dagger\v\A^\dagger\C^\dagger\w\B^\dagger
\\

\stackrel{\is}{=} & 
(\v\D\B\w+(-1)^{b+d}\w\B^\dagger\D^\dagger\v)\A\C \\

& 
-(-1)^{b}((-1)^c \v\D\C^\dagger\w+(-1)^d \w\C\D^\dagger\v)\A\B^\dagger\\

& 
- (-1)^{b+d} \A\w\B^\dagger\D^\dagger\v\C+(-1)^{b+d} \A\w\C\D^\dagger\v\B^\dagger
\\

& 
-(-1)^{a+d} \D^\dagger\v(\B\w\C-(-1)^{b+c}\C^\dagger\w\B^\dagger)\A^\dagger
\\

\stackrel{\is}{=} &
4([\v\D\B\w][\A\C]-(-1)^{b+c}[\v\D\C^\dagger\w][\A\B^\dagger])\\

&
+(-1)^{a+d}(\B\C\D^\dagger\v-\D^\dagger\v\C\B)\w\A^\dagger\\

& +(-1)^{b+d} \A\w(\C\D^\dagger\v\B^\dagger-\B^\dagger\D^\dagger\v\C).
\ea
\label{proof:generalf}
\ee
By (\ref{new:reduction}),
\[\ba{cl}
& (-1)^{d}(\B\C\D^\dagger\v-\D^\dagger\v\C\B)\\

=& (-1)^{d}\B\C\D^\dagger\v-(-1)^{b+c}\v\D\C^\dagger\B^\dagger\bigstrut\\

&
+(-1)^{b+c}(\v\D-(-1)^d \D^\dagger\v)\C^\dagger\B^\dagger\\

&
+(-1)^{c+d}2\,\{\D^\dagger\v(-[\C^\dagger]\B+\C^\dagger[\B])\},
\ea\]
we get from (\ref{proof:generalf}) the following:
\[\ba{cl}
& 4\,{[}\A\v\D][\B\w\C]
\\
\stackrel{\is}{=} &
4([\v\D\B\w][\A\C]-(-1)^{b+c}[\v\D\C^\dagger\w][\A\B^\dagger])\\

&
+4\,[\{(-1)^d[\B\C\D^\dagger\v]+(-1)^{b+c}[\v\D]\C^\dagger\B^\dagger
\\

& 
-(-1)^{c+d}\D^\dagger\v\B[\C^\dagger]+(-1)^{c+d}\D^\dagger\v\C^\dagger[\B]\}
(-1)^a\w\A^\dagger]\\

& \hfill
+(-1)^b \A\w\{-(-1)^c\underline{\v\D\C^\dagger}\B^\dagger+(-1)^c\B^\dagger\C^\dagger\v\D\\

& \hfill
+(-1)^d\underline{\C\D^\dagger\v}\B^\dagger-(-1)^d\B^\dagger\D^\dagger\v\C\}
\\
\ea\]

\[\ba{cl}
\stackrel{\is}{=} &
4([\v\D\B\w][\A\C]-(-1)^{b+c}[\v\D\C^\dagger\w][\A\B^\dagger])\phantom{WWWW}\\

& -4\{
(-1)^d [\A\w][\B\C\D^\dagger\v]+[\A\w\B\C][\v\D]\\

&\hfill
+(-1)^b [\A\w\B^\dagger\v\D][\C]
-[\A\w\C\v\D][\B]\}.
\ea
\]
\endproof

Clearly each term in the result of (\ref{prop:generalf})
has lower leader than the input.
To better understand this reduction formula, we 
write it in tableau form:
\be\ba{lll}
\left[\ba{c} 
\A\v\D\\ 
\B\w\C
\ea\right]

&=& 
\left[\ba{c} 
\A\underline{\w\C}\v\D\\ 
\B
\ea\right]

+\left[\ba{c} 
\A((-1)^{b+1}\underline{\B\w})^\dagger\v\D\\ 
\C
\ea\right]\\

&& \Bigstrut
+\left[\ba{c} 
\v\D\\ 
((-1)^b\B^\dagger)\w((-1)^a\underline{\A})^\dagger((-1)^c\C^\dagger)
\ea\right]\\

&& \phantom{=}\Bigstrut
+\left[\ba{c} 
\underline{\w\C}((-1)^{d+1}\v\D)^\dagger\\ 
\B((-1)^a\underline{\A})^\dagger
\ea\right]\\

&& \phantom{==}\Bigstrut
+\left[\ba{c} 
((-1)^{b+1}\underline{\B\w})^\dagger((-1)^{d+1}\v\D)^\dagger\\ 
\underline{\A}\C
\ea\right]\\

&& \phantom{==-}\Bigstrut
+\left[\ba{c} 
\A\underline{\w}\\ 
((-1)^b\B^\dagger)\underline{\v\D}((-1)^c\C^\dagger)
\ea\right].
\ea
\label{shuffle:explain}
\ee

As $\l_\A\succ \v\succ \w$ and $\l_\A\succ \t_\D$, to decrease the order
of the leader, in the first line of 
(\ref{shuffle:explain}), a subsequence or 
reversed subsequence of the second row of the input tableau is moved up between  
$\A, \v$ of the first row, with the requirement that the subsequence be led by $\w$.  
In the second line of (\ref{shuffle:explain}), $\A$ is moved down behind $\w$ of the second row.

In the third and fourth lines of (\ref{shuffle:explain}), the leading subsequence $\A$ 
of the first row is 
commuted with a subsequence or reversed subsequence of the second row led by $\w$. 
In the last line of (\ref{shuffle:explain}), the trailing subsequence $\v\D$ of the first row is 
commuted with variable $\w$ of the second row.

We make comparison with the shuffle formula for straightening
in classical bracket algebra \cite{sturmfels-white}. 
In the classical bracket algebra over 3D vector space, where the exterior product is also denoted by 
juxtaposition of elements,
suppose that
\[
\left[\ba{ccc}
\a &\v & \d \\
\b &\w & \c
\ea
\right]
\]
is not straight: $\a\v\d$ is ascending, so is $\b\w\c$; $\a\b$ is non-descending, but
$\v\succ \w$. The shuffle formula is obtained as follows.
For any four vectors $\v, \d, \b, \w$
of the 3D vector space, their exterior product equals zero.
By
\be\ba{lll}
0 &=& \a\vee (\v\d\b\w)\vee \c\\
&=& \phantom{-}
[\a\v\d][\b\w\c]-[\a\v\b][\d\w\c]+[\a\v\w][\d\b\c]\\
&&
+[\a\d\b][\v\w\c]-[\a\d\w][\v\b\c]+[\a\b\w][\v\d\c],
\ea
\ee
where ``$\vee$" is the dual of the exterior product called the {\it meet product} 
\cite{white},
we get the following shuffle formula, also called
{\it van der Waerden relation}:
\[\ba{lll}
\left[\ba{ccc}
\a &\hskip -.1cm \v &\hskip -.1cm  \d \\
\b &\hskip -.1cm \w &\hskip -.1cm  \c
\ea
\right] 

&=&
\left[\ba{ccc}
\a &\hskip -.1cm \v &\hskip -.1cm  \underline{\b} \\
\underline{\d} &\hskip -.1cm \w &\hskip -.1cm  \c
\ea
\right]
+
\left[\ba{ccc}
\a &\hskip -.1cm \v &\hskip -.1cm  \underline{\w} \\
\b &\hskip -.1cm \underline{\d} &\hskip -.1cm  \c
\ea
\right]\\
\ea\]

\be\ba{lll}
\phantom{WWWW}
&& \phantom{-}\Bigstrut
+\left[\ba{ccc}
\a &\hskip -.1cm \underline{\b} &\hskip -.1cm  \d \\
\underline{\v} &\hskip -.1cm \w &\hskip -.1cm  \c
\ea
\right]
+\left[\ba{ccc}
\a &\hskip -.1cm \underline{\w} &\hskip -.1cm  \d \\
\b &\hskip -.1cm \underline{\v} &\hskip -.1cm  \c
\ea
\right]\\

&& \phantom{--}\Bigstrut
-\left[\ba{ccc}
\a &\hskip -.1cm \underline{\b\ \ \w} \\
& \hskip -.1cm \underline{\v\ \ \d} &\hskip -.1cm  \c
\ea
\right].
\ea
\label{def:vw}
\ee
(1) The first line commutes $\d$ of the first row and one of the first two vectors of the second row;\\
(2)
the second line commutes $\v$ of the first row and one of the first two vectors of the second row;\\
(3)
the last line commutes $\v\d$ of the first row and the first two vectors of the second row.

For straightening in classical bracket algebra, there are other operations besides (\ref{def:vw}).
When $\a\succ \b$, we only need to commute the two brackets. When $\d\succ \c$ while $\a\preceq \b$
and $\v\preceq \w$, we have another formula for straightening.
By
\[\ba{lll}
0 &=& (\a\v)\vee (\d\b\w\c)\\
&=& \phantom{-} [\a\v\d][\b\w\c]-[\a\v\b][\d\w\c]\\
&& +[\a\v\w][\d\b\c]-[\a\v\c][\d\b\w],
\ea
\]
we get the following {\it Grassmann-Pl\"ucker relation}: 
\[
\left[\ba{ccc}
\a &\hskip -.1cm \v &\hskip -.1cm  \d \\
\b &\hskip -.1cm \w &\hskip -.1cm  \c
\ea
\right]
=
\left[\ba{ccc}
\a &\hskip -.1cm \v &\hskip -.1cm  \underline{\b} \\
\underline{\d} &\hskip -.1cm \w &\hskip -.1cm  \c
\ea
\right]
+
\left[\ba{ccc}
\a &\hskip -.1cm \v &\hskip -.1cm  \underline{\w} \\
\b &\hskip -.1cm \underline{\d} &\hskip -.1cm  \c
\ea
\right]
+
\left[\ba{ccc}
\a &\hskip -.1cm \v &\hskip -.1cm  \underline{\c} \\
\b &\hskip -.1cm \w &\hskip -.1cm  \underline{\d}
\ea
\right].
\]
The last vector $\d$ of the first row commutes in turn with
every vector of the second row.

\section{Conclusion}
\setcounter{equation}{0}
\vskip .2cm

In the bottom-up approach to manipulating brackets, long brackets are expanded into
basic ones by Caianiello expansion, in the end only brackets of length 2 and 3 are left for
further algebraic manipulations. This approach proves to be inefficient in practice, despite the fact
that there are straightening algorithms for polynomials of basic 
invariants \cite{hopf}, \cite{mourrain}.

Uni-bracket polynomials provide a top-down approach to manipulating brackets.
Given a bracket polynomial, by ``ungrading", each bracket but one in every term is expanded 
into a vector-variable binomial, and the bracket polynomial is changed into a uni-bracket one.
Algebraic manipulations of uni-bracket polynomials can take full advantage of the associativity of the
vector-variable product and the symmetries within a uni-bracket. The 
Gr\"obner base ${\cal G}^\square[{\cal M}]$ provided by this paper
further fulfills the arsenal of symbolic manipulations on uni-bracket polynomials.

The last section of this paper suggests a third approach to manipulating brackets by
algebraic manipulations directly upon the input brackets. To establish this approach
there are many research topics ahead: division among bracket polynomials, 
properties of principal ideals, bracket polynomial factorization, and
simplification by reducing the number of terms, {\it etc}.
This seems to be a promising approach.

\vskip .15cm
This paper is supported partially by NSFC 10871195, \\
60821002/F02, and NCMIS of CAS. The Gr\"obner bases for $m=5,6$ were first computed by Dr. L. Huang
on his implementation of the classical non-commutative Gr\"obner base algorithm \cite{mora}.


\begin{thebibliography}{}
\vskip .6cm

\bibitem{altmann} Altmann, S.L. {\it Rotations, Quaternions, and Double Groups}. Oxford University Press, Oxford, 1986.

\bibitem{brini} Bravi, P. and Brini, A. Remarks on invariant geometric calculus, Cayley-Grassmann algebras
and geometric Clifford algebras. In: Crapo, H. and Senato, D. (eds.), {\it Algebraic Combinatorics and
Computer Science}, Springer, Milano, pp. 129-150, 2001.

\bibitem{hopf} DeConcini, C., Eisenbud, E. and Procesi, C. Hodge algebras, {\it Ast\'erisque} {\bf 91},
1982.

\bibitem{hestenes} Hestenes, D. and Sobczyk, G. {\it Clifford Algebra to Geometric Calculus}. D. Reidel, Dordrecht,
Boston, 1984.

\bibitem{li07} Li, H. A recipe for symbolic geometric computing: long geometric product, BREEFS and Clifford factorization.
In: Brown, C.W. (ed.), {\it Proc. ISSAC 2007}, ACM Press, New York, pp. 261-268, 2007.

\bibitem{li} Li, H. {\it Invariant Algebras and Geometric Reasoning}. World Scientific, Singapore, 2008.

\bibitem{li13} Li, H., Huang, L., and Liu, Y. Normalization of quaternionic polynomials.
arxiv: 1301.5338v1 [math.RA], 2013.

\bibitem{lounesto} Lounesto, P. {\it Clifford Algebras and Spinors}. Cambridge University Press, Cambridge, 1997.

\bibitem{mora} Mora, T. An introduction to commutative
and noncommutative Gr\"{o}bner bases.
{\it Theoretical Computer Science} {\bf 134} (1994) 131-173.

\bibitem{mourrain} Mourrain, B. and Stolfi, N. Computational symbolic geometry. In:
White, N. (ed.), {Invariant Methods in Discrete and Computational Geometry}, D. Reidel, Dordrecht,
Boston, pp. 107-139, 1995.

\bibitem{olver} Olver, P.J. {\it Classical Invariant Theory}. Cambridge University Press, Cambridge, 1999.

\bibitem{sturmfels} Sturmfels, B. Computing final polynomials and final syzygies using Buchberger's Gr\"obner basis method.
{\it Result in Math.} {\bf 15}: 551-560, 1989.

\bibitem{sturmfels-white} Sturmfels, B. and White, N.
Gr\"obner bases and invariant theory. {\it Advances in Math.} {\bf 76}: 245-259, 1989.



\bibitem{white} White, N. The bracket ring of combinatorial geometry I. {\it Trans. Amer. Math. Soc.} 
{\bf 202}: 79-103, 1975.

\end{thebibliography}
\end{document}